\documentclass{aa}
\usepackage[utf8]{inputenc}
\usepackage[varg]{txfonts}
\bibpunct{(}{)}{;}{a}{}{,}

\newcommand{\disperse}{DisPerSE}
\newcommand{\dnode}{$d_{\mathrm{CP}}$}
\newcommand{\dfil}{$d_{\mathrm{fil}}$}
\newcommand{\dskel}{$d_{\mathrm{skel}}$}

\title{On the relative effect of nodes and filaments of the cosmic web on the quenching of galaxies and the orientation of their spin}
\author{Nicola Malavasi\inst{1}
\and
Mathieu Langer\inst{1}
\and
Nabila Aghanim\inst{1}
\and
Daniela Gal{\'a}rraga-Espinosa\inst{1}
\and
C{\'e}line Gouin\inst{1}}

\institute{Universit{\'e} Paris-Saclay, CNRS, Institut d'astrophysique spatiale, 91405, Orsay, France}

\date{Received: 1st January 2019 / Accepted: 1st January 2019}

\abstract{Filaments and clusters of the cosmic web have an impact on the properties of galaxies, switching off their star-formation, contributing to the build-up of their stellar mass, and influencing the acquisition of their angular momentum. In this work we make use of the IllustrisTNG simulation, coupled with the \disperse~cosmic web extraction algorithm, to test which is the galaxy property most affected by the cosmic web and, conversely, to assess the differential impact of the various cosmic web features on a given galaxy property. Our aim is to use this information to better understand galaxy evolution and to identify on which galaxy property future efforts should focus to detect the cosmic web from the galaxy distribution. We provide a comprehensive analysis of the relation between galaxy properties and cosmic web features. We also perform extensive tests in which we try to disentangle the effect of local overdensities of galaxies on their properties from the effect of the large scale structure environment. Our results show that star-formation is the quantity that shows the strongest variation with the distances from the cosmic web features, but it is also the one that shows the strongest relation to the local environment of galaxies. On the other hand, the direction of the angular momentum of galaxies is the property that shows the weakest trends with distance from cosmic web features, while also being more independent from the local environment of galaxies. We conclude that the direction of the angular momentum of galaxies and its use to improve our detection of the cosmic web features could be the focus of futures studies benefiting from larger statistical samples.}

\titlerunning{Spin vs. SFR vs. mass of galaxies in filaments and nodes}
\authorrunning{Malavasi et al.}


\begin{document}

\maketitle

\section{Introduction}
Galaxies inhabit a complex network of structures called the large-scale structure of the Universe (LSS) or the cosmic web \citep{deLapparent1986, Bond1996, AragonCalvo2010}. The cosmic web formed through the gravitational collapse of initial fluctuations of the density field \citep{Zeldovich1970collapse, Zeldovich1970tidalfield} with matter departing from under-dense regions (which became voids) and collapsing to form walls (two-dimensional, planar structures which surround voids), filaments (one-dimensional, linear structures found at the intersection of walls) and finally flowing inside filaments to reach clusters, found at the intersection of filaments. Throughout the duration of this process of structure formation, galaxies evolve while flowing within the cosmic web and experiencing strong changes in their environment.

It is now well assessed that the cluster environment affects properties of galaxies such as their mass, star-formation (SFR), colours, and morphology (\citealt{Dressler1980, Dressler1986}; for reviews see e.g. \citealt{BoselliGavazzi2006, BoselliGavazzi2014}). Galaxies in clusters generally have a more elliptical morphology, redder colours, larger mass, and lower SFR with respect to galaxies in less dense environments. With the advent of wide-area spectroscopic surveys our ability to explore the filaments of the cosmic web and their impact on galaxy properties has increased. Analysing data from the Sloan Digital Sky Survey \citep[SDSS,][]{AbazajianSDSS, AlamSDSS}, the VIMOS Public Extragalctic Redshift Survey Multi-$\lambda$ Survey \citep[VIPERS-MLS,][]{Moutard2016II, Moutard2016I, Scodeggio2018}, COSMOS2015 \citep{Laigle2015cosmosphotoz}, the Galaxy And Mass Assembly survey \citep[GAMA,][]{Driver2009}, and the WISExSuperCOSMOS survey \citep[WISExSCOS,][]{Bilicki2016}, \citet{Alpaslan2016, Kuutma2017, Chen2017, Malavasi2017, Laigle2018, Kraljic2018, Bonjean2020, Rost2020} found that massive galaxies and passive galaxies are located closer to the spine of filaments. This effect was also evidenced by \citet{Salerno2020} who also distinguished between galaxies accreting onto clusters isotropically contrasted to those following the filaments direction and found that the quenching of galaxies is enhanced for those that arrive onto clusters following filaments with respect to isotropic accretion \citep[see also e.g.][]{Gouin2020}. Similar trends were observed also in numerical simulations (e.g. by \citealt{Laigle2018, Kraljic2019} using the Horizon-AGN simulation, \citealt{Dubois2014}).

The features of the cosmic web (clusters and filaments) thus both affect galaxy properties. In the case of star formation, interaction with the intra-cluster medium (e.g. ram pressure stripping) or tidal interactions with the cluster potential are known to be quenching mechanisms in clusters \citep[see e.g.][]{DeLucia2007}. With respect to filaments, \citet{Song2020} also invoke an inefficient transfer of gas from the outer parts of the haloes to the inner parts as a possible way to prevent SFR from happening. This inefficient transfer is caused by the alignment of the vorticity of the gas flow in filaments that accretes high-angular momentum gas on the haloes. The interactions of galaxies with both clusters and filaments could also explain the quenching of the SFR by the disconnection of the galaxies from the primordial filaments that supply cold gas to sustain SFR, \citep[an event called ``cosmic web detachment'',][which would essentially produce quenching by starvation]{AragonCalvo2019}. The global combination of all these quenching mechanisms should have the result of producing trends in the SFR of galaxies to decrease with the distance to filaments and clusters.

When the properties of galaxies are analysed as a function of their position with respect to the filaments of the cosmic web, a further effect can be extracted from the data, regarding their angular momentum (i.e. their spin). According to the tidal torque theory \citep{Peebles1969TTT, Doroshkevich1970TTT, White1984TTT, CatelanTheuns1996TTT, Crittenden2001TTT}, dark matter haloes which host galaxies acquire angular momentum due to a torque resulting from a misalignment of their inertia tensor and the external tidal field. Although tidal torque theory generally predicts to a fair degree of accuracy the amplitude and direction of the angular momentum of haloes acquired throughout their linear evolution, later, non-linear effects can significantly modify them \citep{Porciani2002, Dekel2001, Dutton2012}. The tidal torque theory has been later expanded to take into account the fact that the filaments and nodes of the cosmic web both provide and are shaped by the tidal field which defines the direction of the angular momentum of haloes \citep[constrained tidal torque theory,][]{Codis2015, Laigle2015vorticity}. This context also provides an explanation (in terms of the merging of haloes, \citealt{BettFrenk2011, Welker2014}, or smooth secondary accretion, \citealt{Laigle2015vorticity,GaneshaiahVeenaCosmicBalletI}) for the change of spin direction due to non-linear halo evolution spent in the filaments and nodes of the cosmic web. According to constrained tidal torque theory and taking into account subsequent halo evolution, haloes are formed with a spin parallel to the filaments of the cosmic web (in particular low-mass ones) which changes to perpendicular both with increasing halo mass \citep[e.g.][]{AragonCalvo2007spin, Codis2012, Hahn2007evolution, Hahn2007properties} and with the flowing of haloes along the filaments throughout cosmic time (see e.g. \citealt{GaneshaiahVeenaCosmicBalletIII, WangKang2017, Codis2012}, but contrast with \citealt{Trowland2013}). For this reason, we could expect a larger fraction of haloes with their spin perpendicular to filaments close to clusters, as the endpoint of this evolutionary process. At the same time, the spin could also become randomly oriented with respect to the filaments in clusters, as these are essentially regions where the flows from multiple filaments intersect.

The result of this process is a mass transition visible in the distribution of the angles between the spin of haloes and the direction of the closest filament, with low-mass haloes having their spin aligned with the filaments and high mass haloes having their spin perpendicular. Several works have explicitly tried to bracket this mass transition in both simulated and observed data. In particular, for example, \citet{GaneshaiahVeenaCosmicBalletI} explored the alignment of dark matter haloes with respect to filaments in the Planck-Millennium high-resolution N-body simulation \citep{McCullagh2017, Baugh2019}, finding a transition mass for the spin of the whole halo at $M_{DM} =  5 \times 10^{11} M_{\sun}$. Similarly, analysing dark matter haloes in the Horizon $4\pi$ N-body simulation \citep{Teyssier2009}, \citet{Codis2012} detected a larger transition mass ($M_{DM} = 5 \times 10^{12} M_{\sun}$). These results are consistent with \citet{Kraljic2020} who detected a transition mass of $M_{DM} = 10^{12} M_{\sun}$ in the Simba simulation \citep{Dave2019}. When galaxies instead of haloes are considered, trends become weaker (also in light of the fact that a misalignment between the spin of galaxies and the haloes in which they are embedded is possible), but are generally maintained (see e.g. \citealt{Codis2018, GaneshaiahVeenaCosmicBalletI, Kraljic2020, Hahn2010}, but contrast with \citealt{GaneshaiahVeenaCosmicBalletII}). For example, \citet{GaneshaiahVeenaCosmicBalletI} found that the transition mass from aligned to perpendicular spin with respect to filaments becomes smaller if the inner regions of haloes are considered. In actual observations, compatible transition masses are detected for galaxies. For example, \citet{Welker2020} analysed the spin alignment of galaxies in the Sydney-Australian Astronomical Observatory (AAO) Multi-object Integral field spectrograph \citep[SAMI galaxy survey,][]{Croom2012SAMI, Bryant2015SAMI}, with respect to filaments detected in GAMA, finding a transition mass between aligned and perpendicular in the range $M^{\ast} = 10^{10.4} \div 10^{10.9} M_{\sun}$. On the other hand, \citet{Krolewski2019} analysed the spin alignment in the Mapping nearby Galaxies at Apache point survey \citep[MaNGA,][]{Bundy2015MaNGA}, and did not detect any signal, neither for the total population of galaxies, nor when dividing by mass. However, when they performed the same analysis in the Illustris-1 \citep{Vogelsberger2014IllustrisI, Nelson2015IllustrisI} and MassiveBlack-II simulations \citep{Khandai2015}, they detected a transition mass at $M^{\ast} \sim 10^{10} M_{\sun}$ in line with what found also by \citet{Kraljic2020}. When kinematic information about the galaxies is not available, shape can be used as a proxy to infer the direction of the angular momentum. Indeed, \citet{Trujillo2006, Paz2008, Zhang2013, Tempel2013, TempelLibeskind2013, Pahwa2016, Chen2019, Wang2020} among others analysed shape alignments in the Sloan Digital Sky Survey \citep[SDSS,][]{York2000} and in the 2MASS Redshift Survey \citep[2MRS,][]{Huchra2012}, finding a different degree of alignment according to the mass and morphology of galaxies, while \citet{Chen2015} explored alignment in the MassiveBlack-II simulation \citep{Khandai2015}.

Both SFR and the angle between the spin of galaxies and the direction of the closest filament should also show dependencies on the local environment in which galaxies are embedded. In the case of SFR, mergers and high-speed interactions between galaxies (harassment) happening both in clusters and in filaments can contribute to the quenching process \citep[as discussed e.g. in][]{Moutard2018,MoutardMalavasi2020}. Mergers experienced by galaxies while flowing in the filaments towards the clusters are also invoked as a reason for a transition of the spin direction from aligned to perpendicular to the closest filament \citep{BettFrenk2011, Welker2014}. The result of this should be trends for the SFR of galaxies to decrease in high density environments and the spin to become more perpendicular in high density environments.

A complementary approach to studying galaxy properties in relation to environment is to use known relations between the distances of galaxies to structures and their observables to better detect and identify features of the cosmic web. In particular, algorithms have been developed to better detect galaxy clusters using relations between galaxy properties and environment, such as the redMaPPer approach \citep{Rykoff2014}, which makes use of the presence of a defined red sequence in clusters (i.e. the fact that galaxies are redder in denser environments) to improve the detection of galaxy clusters from galaxy surveys. Another example is the work by \citet{Rong2016} which makes use of the average alignment of galaxies to improve the detection of filaments around the Coma cluster \citep[see also e.g.][for a possible use of galaxy alignment to detect the cosmic web at high redshift]{Pandya2019}.

The goal of this work is to explore and consolidate these relations, providing a comprehensive view of the trends of galaxy properties with respect to the various features of the cosmic web. Our aim is to explore how the SFR, mass, and direction of the angular momentum of galaxies vary with respect to the distance to the nodes and filaments of the cosmic web. By analysing which property shows the largest variation with respect to a given feature of the cosmic web we aim to provide a useful indication about which galaxy observable should be targeted by future galaxy surveys with the aim of better detecting the elusive filaments of the cosmic web. To achieve this goal we make use of the cosmic web as detected by \citet{Galarraga2019} using the Discrete Persistent Structure Extractor algorithm \citep[\disperse,][]{Sousbie2011a, Sousbie2011b} in the IllustrisTNG simulation subhalo catalogue \citep{Nelson2019IllustrisTNG}.

This paper is organised as follows: we describe the simulation and the algorithm to detect the cosmic web that we use in Section \ref{data}. We describe the configuration of distances from the cosmic web features that we consider in Section \ref{galdistances} and the properties of galaxies which we follow in Section \ref{galprops}. General results are described in Section \ref{generaltrends}, with further considerations in Section \ref{results_refined}. In Section \ref{conclusions} we summarise our results and draw our conclusions. Throughout this paper we use a \citet{PlanckCollaboration2016} cosmology with $\Omega_{\Lambda} = 0.6911$, $\Omega_{m} = 0.3089$, and $h = H_{0}/100 = 0.6774\: \mathrm{km}\, \mathrm{s}^{-1}\,\mathrm{Mpc}^{-1}$.

\section{Data and method}
\label{data}
For this work we exploit the IllustrisTNG cosmological simulation \citep{NaimanIllustrisTNGpres, PillepichIllustrsTNGpres, NelsonIllustrisTNGpres, MarinacciIllustrisTNGpres, SpringelIllustrisTNGpres, Nelson2019IllustrisTNG}. It has been performed with the moving-mesh code AREPO \citep{Springel2010} and it follows the evolution of dark matter, gas, and stars to $z = 0$, implementing a \citet{PlanckCollaboration2016} cosmology. We use the $z = 0$ snapshot of the TNG300-1 box, with a side of $\sim 300~\mathrm{Mpc}$ and $2500^{3}$ dark matter particles for a resolution of $\sim 4 \times 10^{7} M_{\sun}/h$. Galaxies in this simulation are identified with subhaloes detected by the SUBFIND algorithm \citep{Springel2001, Dolag2009}. We follow the same selection as \citet{Galarraga2019} \citep[based on][]{Nelson2019IllustrisTNG}: we discard all subhaloes flagged by the IllustrisTNG as not reliable ({\tt SubhaloFlag = 0}) and we apply a cut in stellar mass ({\tt SubhaloMassType} for star particles) between $10^{9} \leq M^{\ast}/M_{\sun} < 10^{12}$. The final number of subhaloes in our sample is $275\,818$. Throughout the rest of this paper we will use the terms subhaloes and galaxies interchangeably.

The filaments of the cosmic web in the simulation volume have been detected by \citet{Galarraga2019} using the Discrete Persistence Structure Extractor (\disperse, \citealt{Sousbie2011a, Sousbie2011b}). \disperse~ identifies filaments from the galaxy distribution through the measurement of the gradient of the density field. In our case the density field is measured through the Delaunay Tessellation Field Estimator \citep[DTFE,][]{SchaapvdW2000,vdWSchaap2009} applied to the subhaloes selected above to mimic the galaxy distribution\footnote{While it is true that applying \disperse~to the galaxy distribution provides a different skeleton than the true underlying one which would be obtained by running the algorithm directly on the dark matter particle distribution, characterising the differences between the two is beyond the goal of this work. We refer to \citet{Laigle2018} for an example of such an analysis.}. The density field at the position of a considered galaxy is then smoothed by averaging it with the value measured for all galaxies that share with it an edge of the tetrahedrons of the Delaunay tessellation. When \disperse~ is applied to this density field it identifies critical points, i.e. points where the gradient is zero (maxima, minima, and saddles). Maxima and saddles are connected with filaments of the cosmic web, which follow lines of constant gradient in the density field. Each filament is composed of small segments, the size of the edges of the tetrahedrons of the Delaunay tessellation at each position in space. A persistence cut to $3\sigma$ is applied to remove spurious filaments and critical points due to the Poisson noise of the density distribution. The ensemble of filaments (and critical points) thus constructed (called skeleton) is then smoothed, by averaging the positions of the extrema of each small segment with the positions of the extrema of the two contiguous ones (but keeping the positions of maxima and saddles at the extrema of the filaments unchanged). The skeleton is then broken down and artificial critical points (called bifurcations) are inserted at the positions where several filaments cross. This is done to take into account how \disperse~topologically defines filaments: as consistently connecting maxima to saddles. This would result in some filaments perfectly overlapping for part of their path (see Figure 2 of \citealt{Galarraga2019}) from the same maximum to diffrent saddles. Bifurcations inserted at the point where filaments separate to reach separate saddles after having shared a consistent portion of their path solve this problem and avoid duplicating what should be a single portion of a filament. In total, there are 2999 maxima, 4037 bifurcations, and $15\,220$ filaments in the simulation volume. Note that in this work we define filaments in a slightly different way with respect to \citet{Galarraga2019}, where they were defined as structures detected by \disperse~consistently connecting maxima to saddles. In our case, we define as filaments any structure detected by \disperse~connecting a couple of critical points, whatever their type (e.g. maximum-saddle, maximum-bifurcation, bifurcation-saddle). We choose this approach in order to be more consistent with observations, where galaxies can be found inside filaments connected to both dense clusters (identifiable with the maxima of the density field) and unresolved groups (identifiable with bifurcations). In the following we will refer to maxima and bifurcations generically as ``nodes''. This word is used in an astrophysical sense (with the meaning of peaks of the density field) rather than in a topological one (as in a topological context maxima and bifurcations have different definitions).

\section{Analysis}
In this section we present our analysis of the IllustrisTNG simulated data, namely the measurement of the distances of the subhaloes from the components of the cosmic web and the extraction of the various galaxy properties on which we are going to focus in the rest of the paper.

\subsection{Galaxy distances from the features of the cosmic web}
\label{galdistances}
In terms of distances from cosmic web elements, we chose three that could be related to the various evolutionary paths that a galaxy can take while flowing in the LSS. In particular, when considering variations of galaxy properties as a function of the distance to the axis of the filaments, we can connect recovered trends to the process of galaxies infalling onto filaments from within walls. This process is then followed by galaxies flowing inside filaments to reach clusters, which corresponds to a variation in the galaxies' distance from the nodes following the filaments. On the other hand, galaxies can also directly infall onto clusters isotropically. In our case, this would correspond to a variation in the radial distance of the galaxies from the nodes of the cosmic web. We therefore chose the following distances from the cosmic web elements:

\begin{itemize}
\item Distance from a galaxy to the axis of the closest filament (\dfil): distance from each galaxy to the midpoint of the closest segment belonging to a filament.
\item Distance from a galaxy to the closest node (\dnode): Euclidean distance from a galaxy to the closest maximum or bifurcation. This distance is computed only for galaxies that have \dfil$> 1$ Mpc as these galaxies can be considered as being outside the core of the filaments, based on the density profile for filaments derived by \citet{Galarraga2019}.
\item Distance from a galaxy to the node connected to the closest filament following the filaments (\dskel): for each galaxy we consider the closest filament and we consider the distance from the point of the filament closest to the galaxy (i.e. the projection of the galaxy position on the filament) to one of the two critical points connected to that filament. The critical point is consistently chosen as the densest of the two: in case of a filament connecting a maximum and a bifurcation or saddle we choose the maximum, in case of a filament connecting a bifurcation and a saddle we choose the bifurcation, in case of a filament connecting two bifurcartions or two saddles we choose the densest of the considered critical points. This quantity is only computed for galaxies which are located inside the core of the filaments (i.e. having \dfil$< 1$ Mpc).
\end{itemize}

A value of 1 Mpc for the size of the filament core is roughly four times the best-fit scale radius for the Generalised Navarro, Frenk, and White profile \citep{Hernquist1990, Navarro1997, Nagai2007, Arnaud2010} found by \citet[][see their Table 5, top row]{Galarraga2019}. We therefore consider galaxies with \dfil$> 1$ to be situated well outside the filament core. We do note however that this value is significantly lower than the value of 27 Mpc mentioned by \citet{Galarraga2019} as an extreme limit for the filaments density profile. We chose a threshold of \dfil$= 1$ Mpc as a compromise between the best-fit scale radius and the extreme limit of filaments derived in \citet{Galarraga2019}, in order to not be affected by too small number counts in either the in-filament or the out-filament sample, which may have an impact on our conclusions. We also tested what happens to the distributions of galaxy properties as a function of distance from the cosmic web features (shown in Figures \ref{sfrmassangledist} and \ref{distquantitiessamedist}, discussed below) when we consider a threshold distance to identify galaxies inside or outside filaments of \dfil$= 0.25, 0.5, 1, 10, 27$ Mpc. Our conclusions do not change, regardless of the chosen threshold.

\begin{figure}
\centering
\includegraphics[width = \linewidth, trim = 7cm 4cm 7cm 4cm, clip = true]{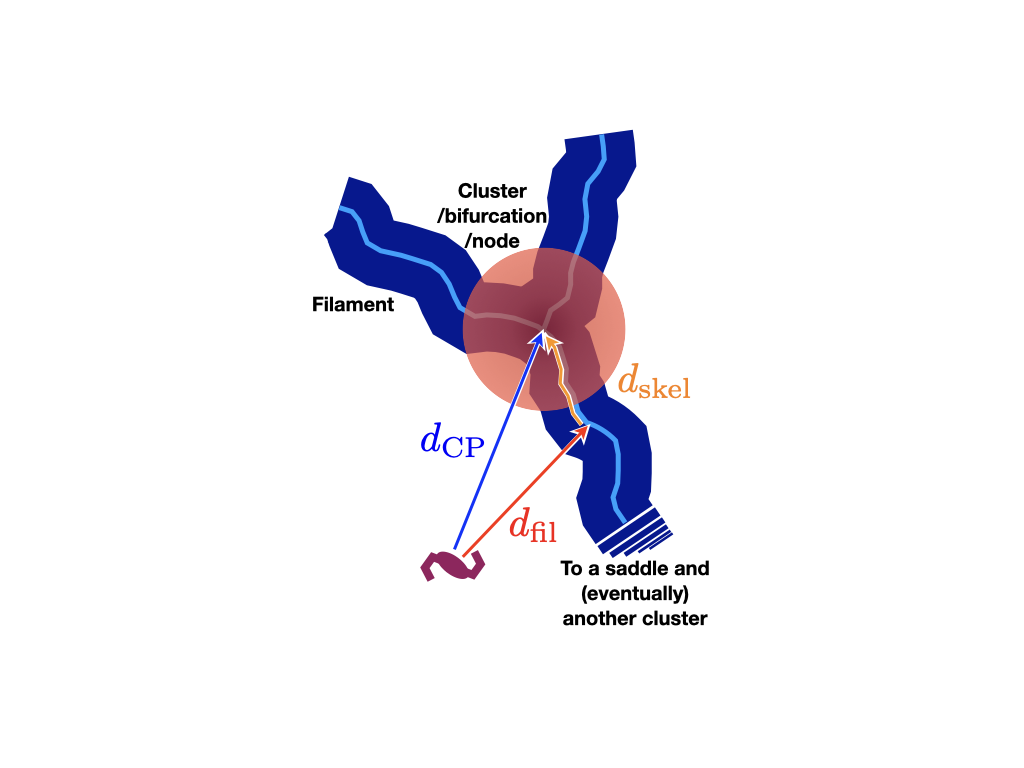}
\caption{Schematic representation of the three distances considered in this study: \dfil, \dnode, and \dskel. \dfil~is the distance from a galaxy to the axis of the closest filament, \dnode~is the Euclidean distance from a galaxy to the closest maximum or bifurcation (computed only for galaxies that have \dfil$> 1$ Mpc), and \dskel~is the distance from a galaxy to the node connected to the closest filament following the filaments (computed only for galaxies that have \dfil$<1$ Mpc).}
\label{schema_cod}
\end{figure}

A schematic view of the considered distances from the cosmic web features is given in Figure \ref{schema_cod}. In the rest of this paper we will refer to the nodes of the cosmic web (maxima and bifurcations) and clusters interchangeably. However we stress here that it is hard to find a direct correspondence between critical points as identified by \disperse~and galaxy groups or clusters \citep[see e.g. Figure 3 of][]{Malavasi2020a}. However, appendix \ref{appendix_thresh_cpdens} shows that considering only the densest critical points in our analysis (i.e. those who are the more likely to match with the position of groups and clusters), has only minor effects on our results.

We stress that the node chosen to compute \dnode~(the closest one to the considered galaxy) may not be the same as the node considered to measure \dskel~(the densest one at the end of the closest filament). Indeed, for about 30 \% of the galaxies in our sample the two do not match. Moreover, for 22 \% of our galaxies the closest node is not connected to the closest filament. We tested the effect of this discrepancy on our results by eliminating from our sample the galaxies for which the closest node and the densest one at the end of the closest filament used to measure \dskel~do not match and re-derived the distributions shown in Figures \ref{sfrmassangledist} and \ref{distquantitiessamedist}. We find no difference in our conclusions when we eliminate these galaxies from our sample.

\subsection{Galaxy properties}
\label{galprops}
As mentioned in the introduction, for this analysis we are going to focus on the three main types of galaxy properties which are known to vary in response to the environment of the cosmic web: galaxy stellar mass, observables related to the amount of star-formation in galaxies (e.g. star-formation rate, fraction of quenched galaxies), and observables related to the relation between the direction of the galaxies' angular momentum and the direction of the closest filament (e.g. angle between spin and filament direction, fractions of galaxies with a parallel or perpendicular alignment between spin and filaments). Both star-formation and angular momentum acquisition in galaxies are correlated with mass. In the case of star-formation, massive galaxies are known to be less star-forming \citep[see e.g.][and references therein]{Strateva2001, BlantonMoustakas2009, Bolzonella2010, Pozzetti2010, Ilbert2013, Malavasi2017Ultravista} while the orientation of the spin with respect to the direction of the large-scale structure changes at the already mentioned transition mass \citep{GaneshaiahVeenaCosmicBalletI, Codis2012, Krolewski2019, Kraljic2020, Welker2020}. For this reason, in the following we will also focus on galaxy quantities which are derived taking galaxy stellar mass into account (e.g. specific star-formation rate) or we will separate the galaxy population into high- and low-mass objects, based on the position of the spin transition mass in our data. Several of these galaxy properties are readily available in the IllustrisTNG subhalo catalogue, we explain in the paragraphs below how we extract them and compute those that are not already present.

\subsubsection{Mass and star-formation}
The stellar mass ($M^{\ast}$) in the subhalo catalogue of IllustrisTNG is defined using the \texttt{SubhaloMassType} field for star particles, cut between $10^{9}$ and $10^{12} M_{\sun}$ following \citet{Galarraga2019}. The subhalo mass distribution presents a large number of low-mass haloes and a progressively decreasing number of high-mass haloes. The break between the two is located at $\sim 3 \cdot 10^{10} M_{\sun}$. To perform our analysis we identify two mass regimes, namely high-mass and low-mass subhaloes. We define high-mass subhaloes as those with $M^{\ast} \geq 10^{11} M_{\sun}$ and low-mass subhaloes as those with $M^{\ast} \leq 10^{10} M_{\sun}$. These thresholds have been chosen as the limits encompassing the stellar mass range where the transition in galaxy spin alignment from parallel to perpendicular is found both in the literature \citep[see e.g.][]{Codis2018,Welker2020} and in the IllustrisTNG subhalo sample analysed here (see Section \ref{angmomdir} and bottom panel of Figure \ref{thetamassdist}). These two values also bracket the position of the knee of the mass function, which represents the typical mass of the average population of galaxies in our sample, and allow us to explore the extreme tails of the mass distribution, therefore increasing the chance of detecting differences in the behaviour of high- and low-mass galaxies.


The Star-Formation Rate (SFR) is defined as the sum of the individual SFRs of all gas cells in a given subhalo (in $M_{\sun}/yr$). Several subhaloes are present with a SFR of zero, due to the fact that no star-forming gas cells were found inside them. We do not eliminate these haloes from the sample, but rather we consider them as belonging to the quenched halo population. As an additional quantity to trace star-formation in a way which takes galaxy stellar mass into account we also focus on the specific SFR (sSFR) defined as $\mathrm{sSFR} = \mathrm{SFR}/M^{\ast}$ and we rely on this quantity (rather than on SFR) to define our quenched galaxy population. We identify quenched galaxies as those with $\mathrm{sSFR} \leq 10^{-11} \mathrm{yr}^{-1}$ \citep[see e.g.][and references therein]{Pozzetti2010, Davidzon2016, Donnari2020}.


\subsubsection{Spin}
\label{galpropsspin}
In the IllustrisTNG halo catalogue, the spin per unit mass ($\boldsymbol{j}$) is defined through the components along each axis of the mass weighted sum of the relative coordinates times the relative velocities of particles. Considering a subhalo and all its  member dark matter particles and gas cells of relative coordinates $\mathbf{r}_{p} = \mathbf{r} - \mathbf{r}_{H}$ (with $\mathbf{r}$ and $\mathbf{r}_{H}$ the absolute coordinates of the considered particle and of the centre of the subhalo to which the particle belongs, respectively) and relative velocities $\mathbf{v}_{p} = \mathbf{v} - \mathbf{v}_{H}$ (with $\mathbf{v}$ and $\mathbf{v}_{H}$ the absolute velocities of the considered particle and of the centre of the subhalo to which the particle belongs, respectively) the spin is defined as

\begin{equation}\label{spineq}
\boldsymbol{j} = \frac{\sum_{p \in H} m_{p} \mathbf{r}_{p} \times \mathbf{v}_{p}}{\sum_{p \in H} m_{p}}
\end{equation}
(in $\mathrm{Mpc} \cdot \frac{km}{s}$).

The index $p \in H$ indicates all particles of a certain type (and with mass $m_{p}$) which belong to the considered subhalo. Unless otherwise stated, in the following we will consider the spin computed using all particles and cells (dark matter, gas, stars, and black holes) belonging to a subhalo (which we refer to as $\boldsymbol{j}_{tot}$). We also compute the spin for individually selected baryonic components such as gas and stars ($\boldsymbol{j}_{gas}$, $\boldsymbol{j}_{stars}$) as well as for the total baryonic component of each subhalo ($\boldsymbol{j}_{gas+stars}$). Results derived specifically with the spin measured using only certain subhalo components are shown in Appendix \ref{appendix_spin_components}. We made this further check to understand whether the relation between the spin direction and the direction of the filaments that we detect when considering the full subhalo is maintained also when considering components more similar to what is targeted with observations of galaxies. The detection of an alignment between filaments and the spin of galaxies is more uncertain in the literature, however Appendix \ref{appendix_spin_components} shows that the trends we detect with the full subhalo are maintained when we consider only the gaseous component. We therefore assume our conclusions to be valid to some extent also for galaxies.

From the spin it is possible to compute the so-called Bullock parameter \citep[][also called {\it spin parameter}]{Bullock2001}. It is defined as

\begin{equation}\label{bullparam}
\lambda = \frac{j}{\sqrt{2} R_{\mathrm{tot}} V_{c, \mathrm{tot}}}
\end{equation}

\noindent where $R_{\mathrm{tot}}$ is the radius which includes all the mass of the subhalo (i.e. the distance from the centre of the subhalo to the farthest particle of any type which is bound to the subhalo), $V_{c, \mathrm{tot}} = \sqrt{GM_{\mathrm{tot}}/R_{\mathrm{tot}}}$ is the circular velocity at $R_{\mathrm{tot}}$, and $M_{\mathrm{tot}} = \sum_{p \in H} m_{p}$. Appendix \ref{appendix_bullock_r500} shows the results obtained when the Bullock parameter is derived using quantities computed at $R_{200}$ instead of $R_{\mathrm{tot}}$.

\begin{figure}
\centering
\includegraphics[width = \linewidth]{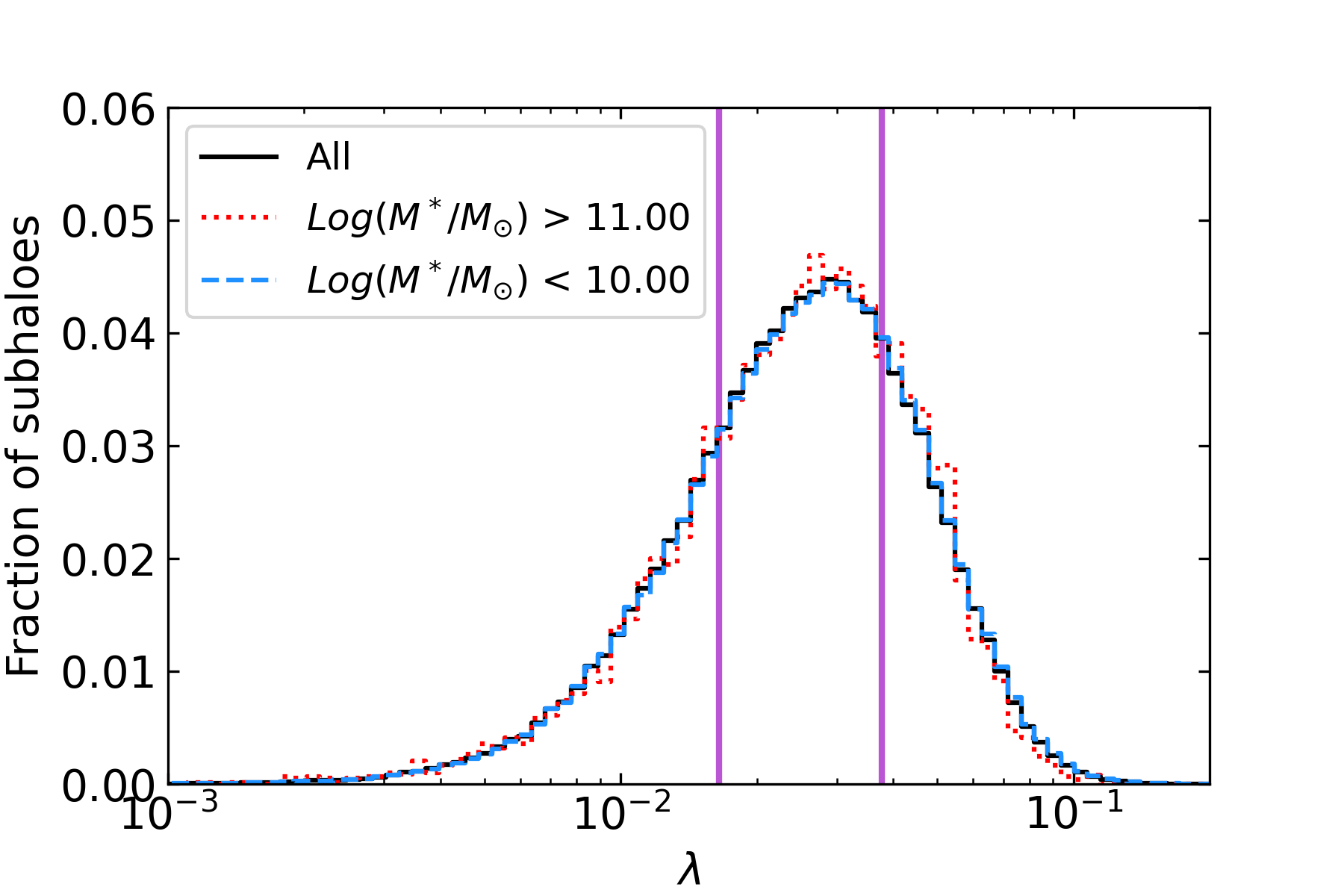}
\caption{Distributions of the Bullock parameter $\lambda$ as derived in Equation \eqref{bullparam}. The black line refers to the total sample, the red dotted line refers to high-mass galaxies and the blue dashed line refers to low-mass galaxies. The vertical purple lines are located at the 25th and 75th percentiles of the $\lambda$ distribution and represent a qualitative distinction between high-spin parameter and low-spin parameter galaxies.}
\label{spingeneraldist}
\end{figure}

Figure \ref{spingeneraldist} shows the distribution of the Bullock parameter $\lambda$ for all the subhaloes in the sample. The Bullock parameter distribution for the total subhalo population is in very good agreement with what derived for other numerical simulations \citep[see e.g.][]{GaneshaiahVeenaCosmicBalletI, Hellwing2021}. The distribution is rather smooth, peaking at a value of $\lambda \sim 3 \cdot 10^{-2}$ and compact, with short tails at higher and lower values of $\lambda$. This figure also shows the distribution of Bullock parameters for high-mass and low-mass galaxies. The two populations do not seem to differ in the distributions of the parameter $\lambda$. In the following, we divide our galaxy population in two samples, namely high-spin parameter galaxies (those with $\lambda$ greater than the 75th percentile of the distribution, i.e. $\lambda = 0.038$) and low-spin parameter galaxies (those with $\lambda$ lower than the 25th percentile of the distribution, i.e. $\lambda = 0.016$). We used the percentiles of the distribution to separate between high- and low-spin parameter galaxies to obtain the behaviour of extreme populations, in terms of spin parameter amplitude.

\subsubsection{Angular momentum vector direction in relation to the LSS}
\label{angmomdir}
For each subhalo in the total sample, we measure the angle $\theta$ between the direction of the angular momentum (spin) vector and the local direction of the filament closest to the considered subhalo (i.e. the direction of the closest segment as output by \disperse). The computation of this angle is made in 3D and when the spin is not perpendicular to the direction of the filament ($\theta \sim 90 \deg$), it can either be aligned ($\theta \sim 0 \deg$) or anti-aligned (i.e. $\theta \sim 180 \deg$). In this work we consider alignment and anti-alignment between filaments and spin as the same situation, for this reason we limit the angle between spin and filaments to the range $\theta \in [0,90] \deg$. In the following we will consider subhaloes as having a spin parallel to the direction of the closest filament if $\theta \leq 30 \deg$, perpendicular if $\theta \geq 60 \deg$, and as having no preferential orientation with respect to the direction of the closest filament if $30 \deg \leq \theta \leq 60 \deg$. We also consider the quantity $\cos(4\theta)$ to separate between the population of subhaloes with spin either parallel or perpendicular to the filaments and the population of subhaloes with no relation between spin and filament direction. In particular, given its period, the quantity $\cos(4\theta)$ takes positive values when $\theta \leq 30 \deg$ or $\theta \geq 60 \deg$ and negative ones when $30 \deg \leq \theta \leq 60 \deg$. In the following we will refer to the case when $\cos(4\theta) > 0$ as subhaloes having an {\it ordered} relation between spin and filament direction.

The expected distribution of values of angle between the direction of the spin and the direction of the closest filament ($\theta$) for a population of galaxies randomly oriented with respect to the surrounding LSS is not uniform in 3D. Indeed, the expected distribution of $\theta$ values for a population of random galaxies is

\begin{equation}\label{ptheta}
P(\theta) = \frac{\sin(\theta)}{2}
\end{equation}

\begin{figure}
\centering
\includegraphics[width = \linewidth]{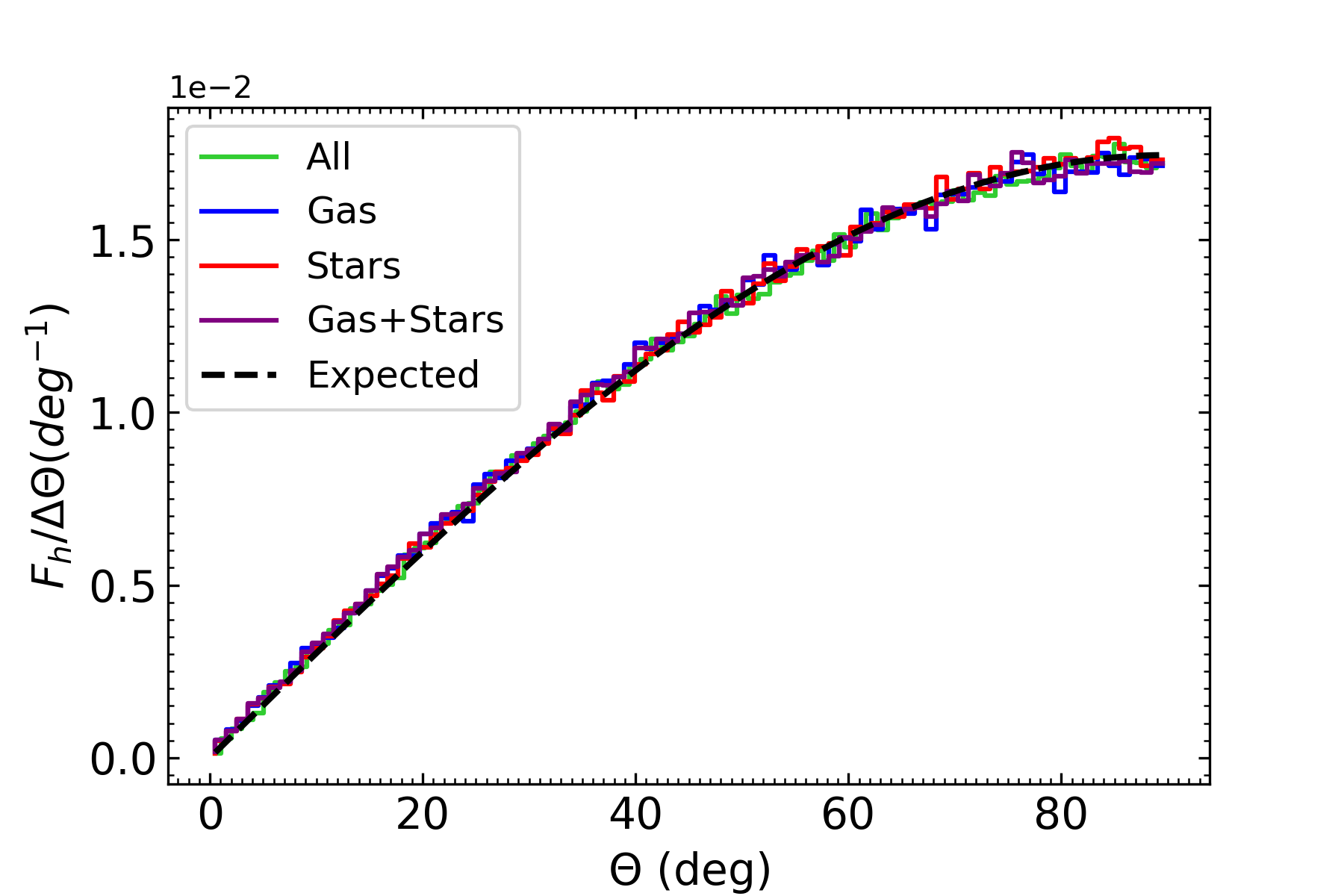}
\includegraphics[width = \linewidth]{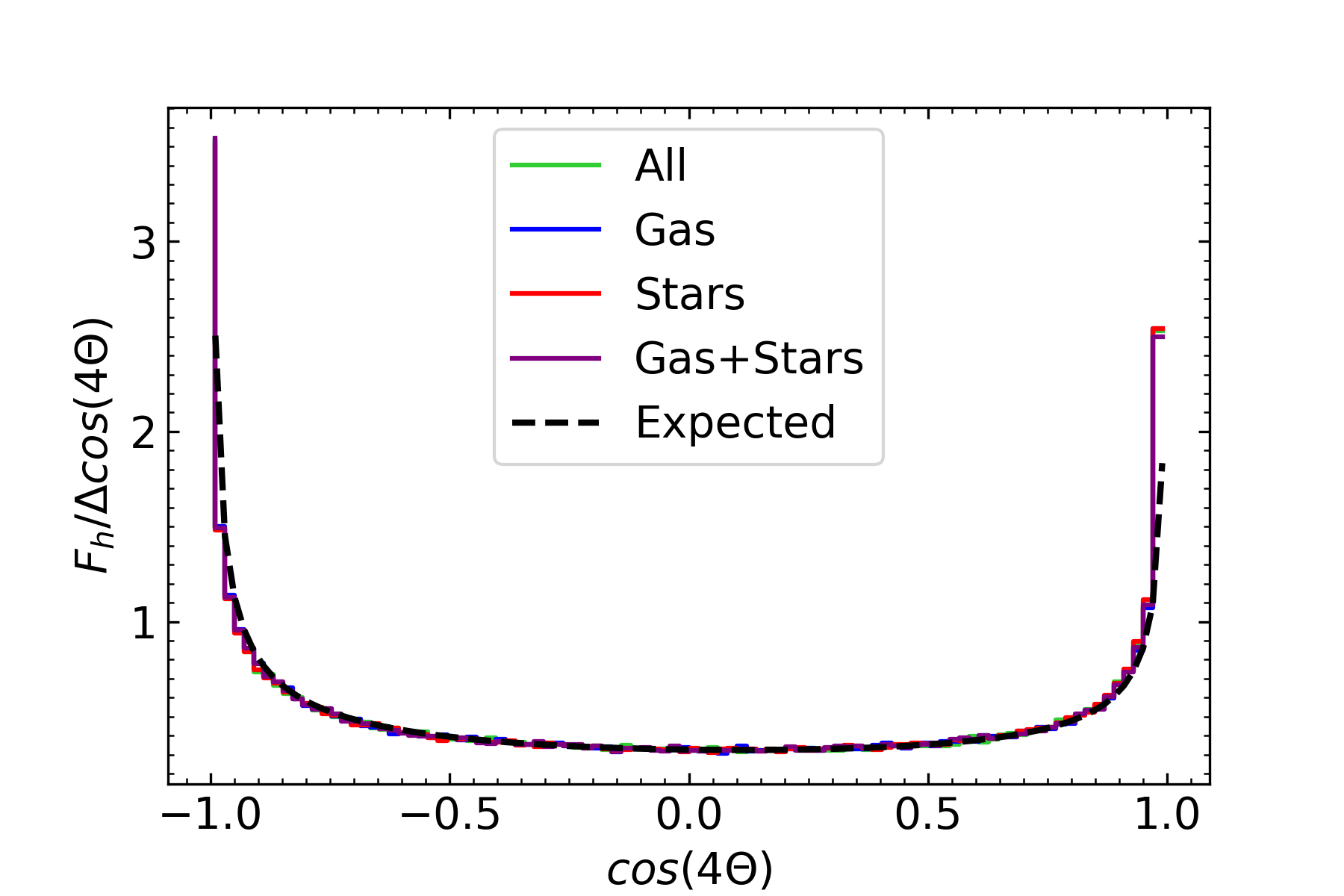}
\caption{Top panel: distribution of the values of the angle between the spin of the subhaloes and the direction of the closest filament ($\theta$). Bottom panel: distribution of values of $\cos(4\theta)$ for all the subhaloes in the box. In both panels, solid lines show the measured distribution of values of $\theta$ and $\cos(4\theta)$ (green: $\boldsymbol{j}_{tot}$, blue: $\boldsymbol{j}_{gas}$, red: $\boldsymbol{j}_{stars}$, purple: $\boldsymbol{j}_{gas+stars}$). The black dashed lines show the expected distribution for a random population of subhaloes (equation \eqref{ptheta}, top panel and equation \eqref{pcos4t}, bottom panel).}
\label{thetageneraldist}
\end{figure}

Figure \ref{thetageneraldist} (top panel) shows how the distribution of angles $\theta$ for the total halo population in the IllustrisTNG box follows the expected distribution for galaxies with random alignments, without particular features regardless of the component used to measure the spin. The distribution of angles varies smoothly between $0 \deg$ and $90 \deg$. The bottom panel of this figure shows the distribution of $\cos(4\theta)$ for the general galaxy population. The expected distribution of values of $\cos(4\theta)$ can be derived from equation \eqref{ptheta} by performing a change of variable to $z = \cos(4\theta)$. The resulting distribution is

\begin{equation}\label{pcos4t}
P(z) = \frac{\sin(\arccos(z)/4)+\sin(\arccos(z)/4+\frac{\pi}{2})}{4\sqrt{1-z^2}}
\end{equation}

Also for this quantity, the observed distribution follows the expected one for a galaxy population with random alignments between the spin of the galaxies and the direction of the closest filament when all the subhaloes in the box are considered.

\begin{figure}
\centering
\includegraphics[width = \linewidth]{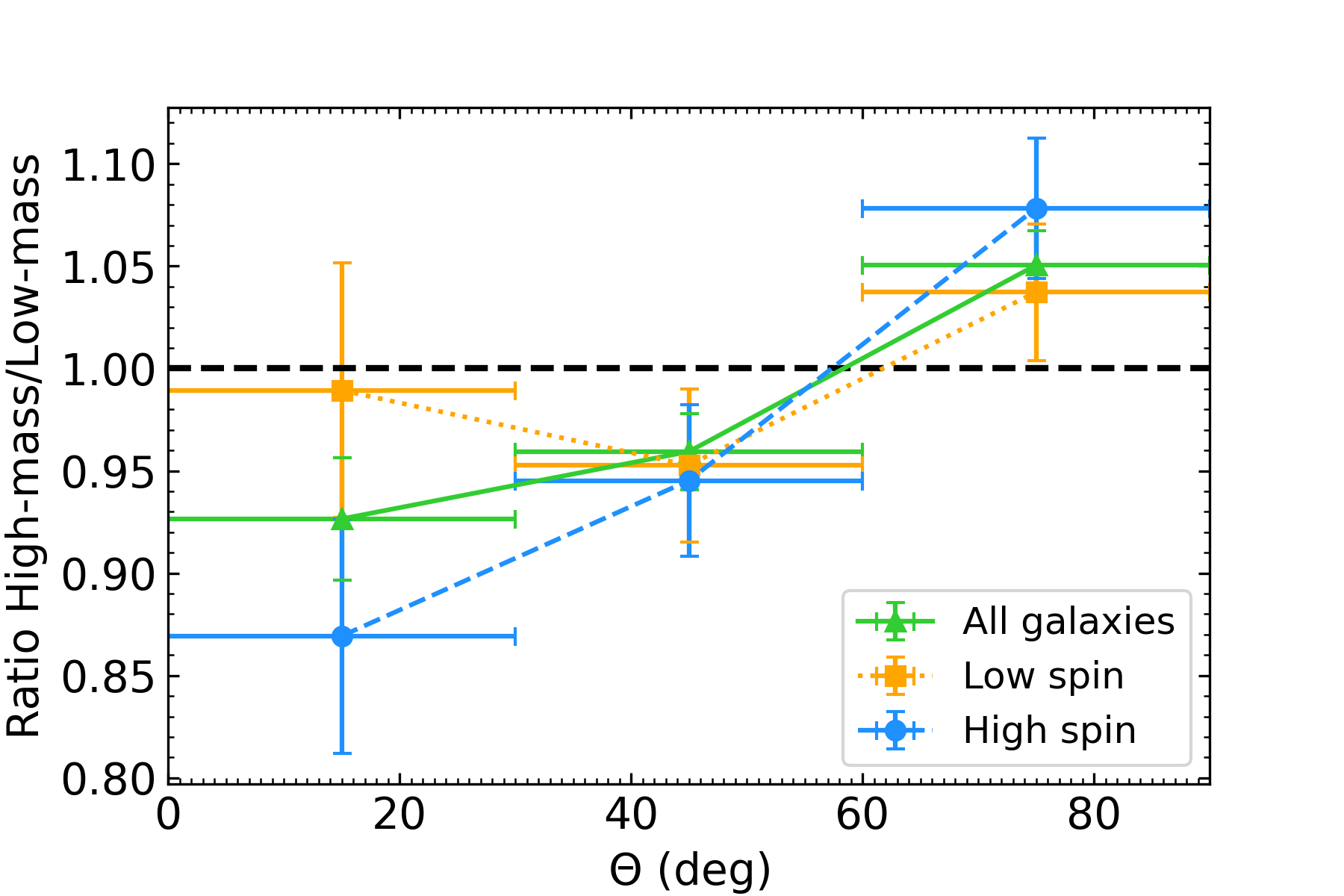}
\includegraphics[width = \linewidth]{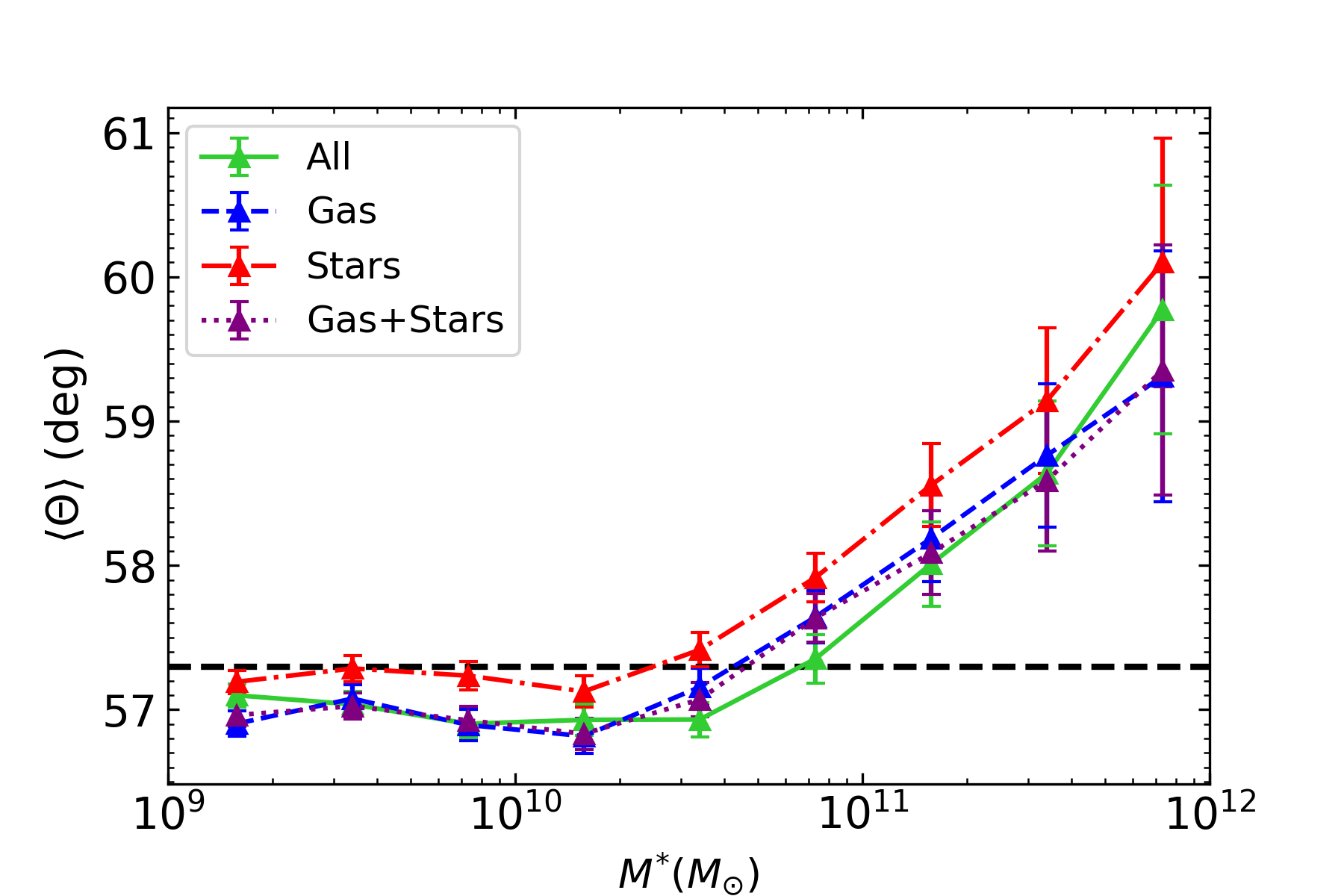}
\caption{Top panel: ratio of the distribution of $\theta$ values (binned in three bins corresponding to parallel and perpendicular orientations and to no preferential orientation), as derived for the high-mass and the low-mass galaxies (only $\boldsymbol{j}_{tot}$ is considered). The green line shows the ratio of high-mass to low-mass $\theta$ distributions for all the galaxies, the orange dotted line for low-spin parameter galaxies, and the cyan dashed line for high-spin parameter galaxies as defined with the percentiles of the Bullock parameter distribution. Bottom panel: the average angle $\theta$ computed in bins of stellar mass. The black dashed line is $\bar{\theta}$ given by equation \eqref{ptheta} for the total subhalo sample. The coloured lines are the distributions for the spin measured with different components (solid green: $\boldsymbol{j}_{tot}$, dashed blue: $\boldsymbol{j}_{gas}$, dot-dashed red: $\boldsymbol{j}_{stars}$, dotted purple: $\boldsymbol{j}_{gas+stars}$).}
\label{thetamassdist}
\end{figure}

Although the distribution of $\theta$ values follows the expected one for a sample of randomly oriented subhaloes, a trend for galaxies to be preferentially parallel (perpendicular) to filaments emerges when low-mass (high-mass) galaxies are selected. In the top panel of Figure \ref{thetamassdist} we present the ratio of the number of high-mass to low-mass galaxies in three bins of $\theta$, corresponding to the cases of parallel, perpendicular, and no preferential orientation of the spin with respect to the filaments. For the general population of the galaxies in the box, there is a clear deficit of high-mass galaxies with their spin direction aligned with the direction of the closest filament. An excess of high-mass galaxies with their spin perpendicular to filaments is also visible. When we split the galaxy population between high-spin parameter and low-spin parameter galaxies using the percentiles of the Bullock parameter distribution, Figure \ref{thetamassdist} shows how in the case of high-spin parameter galaxies the deficit and excess of high-mass galaxies with their spin aligned and perpendicular to the filaments are more significant. In the case of low-spin parameter galaxies, the ratio of the high-mass and the low-mass distribution is consistent with being one.

We explored the mass at which a transition between aligned and perpendicular spin happens in our data. As stated in the introduction, the result of the evolution of galaxies while they flow in the filaments of the cosmic web is a change in the alignment of their spin with the filaments and an increase in stellar mass. This results in low-mass galaxies retaining a spin parallel to the filaments and high-mass galaxies having their spin perpendicular to the filaments. Several works in the literature have tried to bracket the spin transition mass as a mean to shed light onto the process of galaxy evolution in the cosmic web, but although a broad consensus has been reached, no precise mass value has been obtained.

The bottom panel of Figure \ref{thetamassdist} shows the average angle $\langle \theta \rangle$ in a series of increasing mass bins. When all particle types are considered when measuring the spin ($\boldsymbol{j}_{tot}$), the transition between aligned (i.e. $\langle \theta \rangle < \bar{\theta}$, the expectation value for a random sample of haloes distributed following equation \eqref{ptheta}) and perpendicular (i.e. $\langle \theta \rangle > \bar{\theta}$) happens at a mass of $\sim 8 \cdot 10^{10} M_{\sun}$. When only baryonic components are considered, the transition mass decreases to $\sim 4 \cdot 10^{10} M_{\sun}$ (for $\boldsymbol{j}_{gas}$ and $\boldsymbol{j}_{gas+stars}$) and $\sim 2 \cdot 10^{10} M_{\sun}$ (for $\boldsymbol{j}_{stars}$). This range of spin transition masses is in agreement with other works in the literature and represents the first such estimate for the IllustrisTNG simulation. Our choice of limits $10^{10} M_{\sun}$ and $10^{11} M_{\sun}$ to distinguish low- and high-mass galaxies brackets the mass region where the spin transition happens. In particular, while the spin transition mass of $M^{\ast} = 10^{11} M_{\sun}$ detected in our sample sets our high-mass limit, the value we chose for a low-mass limit of $M^{\ast} = 10^{10} M_{\sun}$ is in agreement with what reported by \citet{Codis2018} as a spin transition mass. If we limit our galaxy sample only to high-spin parameter subhaloes, the transition mass value for spin alignment is preserved, while no transition in spin alignment from parallel to perpendicular is visible at any mass for low-spin parameter galaxies (not shown here). This could be an indication of the fact that the direction of the angular momentum vector is better defined for galaxies that have a more prominent spin parameter, therefore leading to a smaller uncertainty in the measurement of the angle between spin and filaments and an increase in the alignment signal that can be extracted for high-spin parameter galaxies.

\subsection{Goal of the analysis}
\begin{figure*}
\centering
\includegraphics[scale = 0.91, trim = 1.25cm 2.5cm 1.25cm 2.5cm, clip = true]{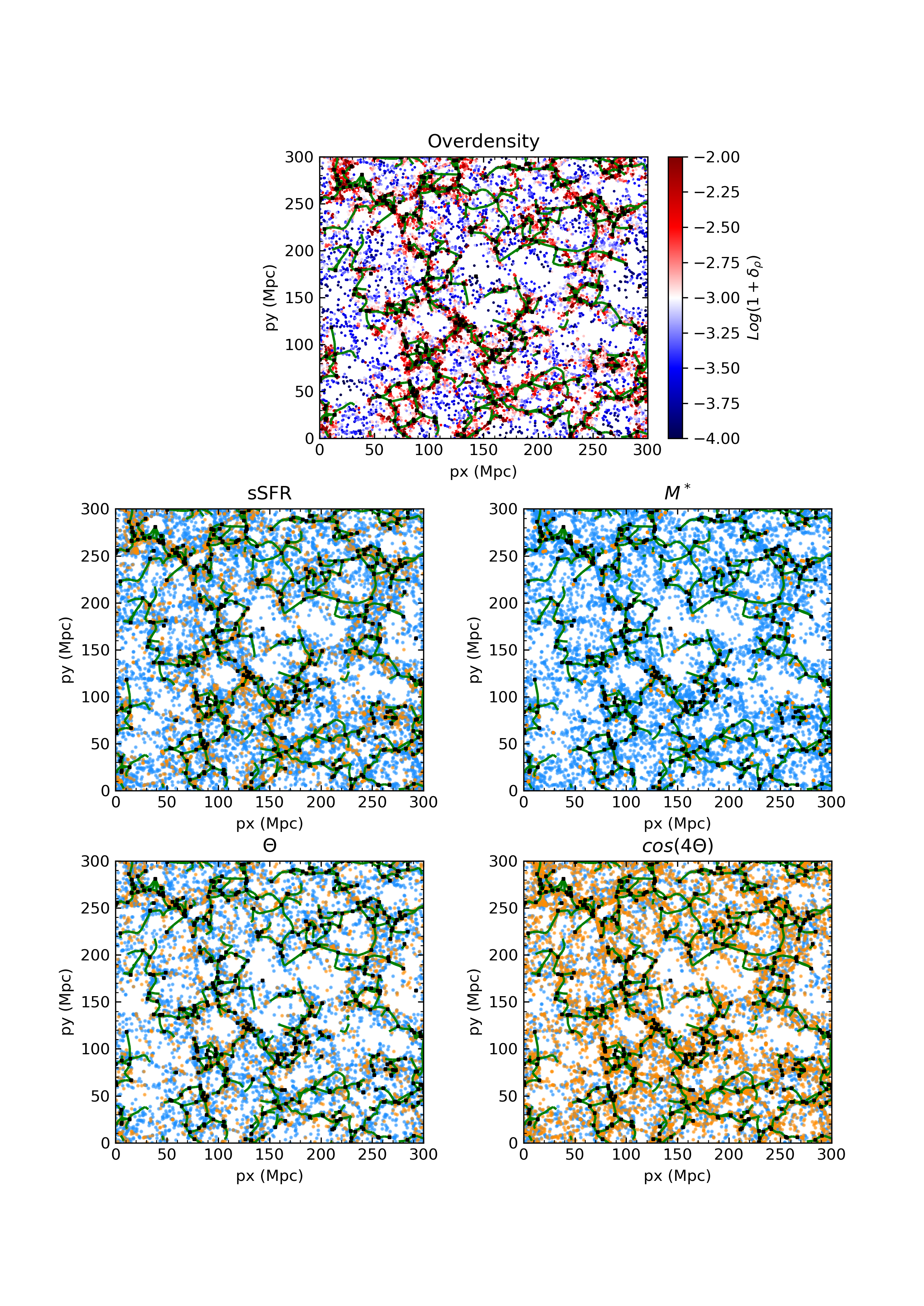}
\caption{Galaxy positions with respect to the cosmic web features. This figure shows a slice 25 Mpc thick of the simulation box, encompassing the most massive subhalo of the simulation. In all panels, points represent galaxies, green lines filaments, and black squares maxima and bifurcations as derived by \disperse. In the top panel, subhaloes are colour-coded according to their local density ($\log(1+\delta_{\rho})$), all subhaloes are shown. In the bottom four panels, subhaloes are colour coded according to which population they belong: star-forming (light blue) or quenched (orange) in the middle left, low-mass (light blue) or high-mass (orange) in the middle right, perpendicular (light blue) or parallel (orange) in the bottom left, no orientation (light blue) or ordered (orange) in the bottom right. See Section \ref{galprops} for the distinction in the various subsets. In the top left and bottom right panels all subhaloes are shown, while in the other panels subhaloes not beloging to the considered populations are not shown.}
\label{boxslices}
\end{figure*}

Figure \ref{boxslices} shows in a qualitative way the trends we explore in the rest of the paper. This figure shows a 25 Mpc thick slice of the simulation box, encompassing the most massive subhalo of the catalogue. In the top panel, the subhaloes in the slice are colour-coded according to the local density contrast as derived directly from the DTFE density ($1+\delta_{\rho} = \rho_{\mathrm{DTFE}}/\langle \rho_{\mathrm{DTFE}} \rangle$, where the average is computed over the full box). Filaments from \disperse~ are overlaid in green and their path precisely follows the density field, as expected. Critical points identified by \disperse~(only maxima and bifurcations are shown for the sake of clarity, shown in black in the figure) are found at the intersection of filaments.  

The four bottom panels show the same filaments and critical points as the top one, but different sets of galaxies are represented. In particular, the figure shows how star-forming and quenched galaxies (with the distinction between the two set at $\mathrm{sSFR} = 10^{-11} \mathrm{yr}^{-1}$), high-mass and low-mass galaxies (with the distinction set at $M^{\ast} \geq 10^{11} M_{\sun}$ and $M^{\ast} \leq 10^{10} M_{\sun}$, respectively), parallel and perpendicular (i.e. $\theta \in [0,30] \deg$ and $\theta \in [60,90] \deg$, respectively), and ordered and without a preferential direction (with the distinction between the two being set at $\cos(4\theta) = 0$) are distributed with respect to the nodes and the filaments. This figure shows that massive galaxies are very rare and mostly located in dense regions at the intersection of filaments, tracing the density peaks. On the other hand, low mass galaxies are more uniformly distributed around filaments. The same is true for star-forming galaxies, while quenched ones mostly tend to cluster at nodes, with a few around the filaments. However, parallel galaxies do not show a particular tendency for clustering around nodes. Rather, they tend to be more uniformly distributed around filaments, highlighting the position of the features of the cosmic web. This is even more evident in the case of ordered galaxies, which clearly delineate the position of both high-density and low-density structures of the cosmic web (nodes and filaments). Although qualitatively, this figure shows how different properties of galaxies may trace differently the various features of the cosmic web. Due to the fact that galaxy and halo spin direction is mainly initially set by the filaments (constrained tidal torque theory) and changed by subsequent evolution of the galaxies in the filaments while they flow towards the clusters, we expect the alignment to be stronger in the filament environment, which therefore we expect to be better traced by this galaxy property. On the other hand, other properties such as SFR are strongly affected e.g. by the dense and hot gaseous environment of clusters and should therefore be a better tracer of nodes. The goal of this work is to test whether these expectations are correct. While the effect of the clusters on the SFR and the mass of galaxies has been extensively tested and recent results started to investigate the effect of filaments on both these quantities and the spin alignment of galaxies, in this work we try to quantify the relative importance of mass, SFR, and spin to characterise the cosmic web, and what is the impact of clusters and filaments on these observables.

\section{Results}
\label{generaltrends}
\begin{figure*}
\centering
\includegraphics[trim = 2cm 1cm 1cm 0cm, clip = true, width = \linewidth]{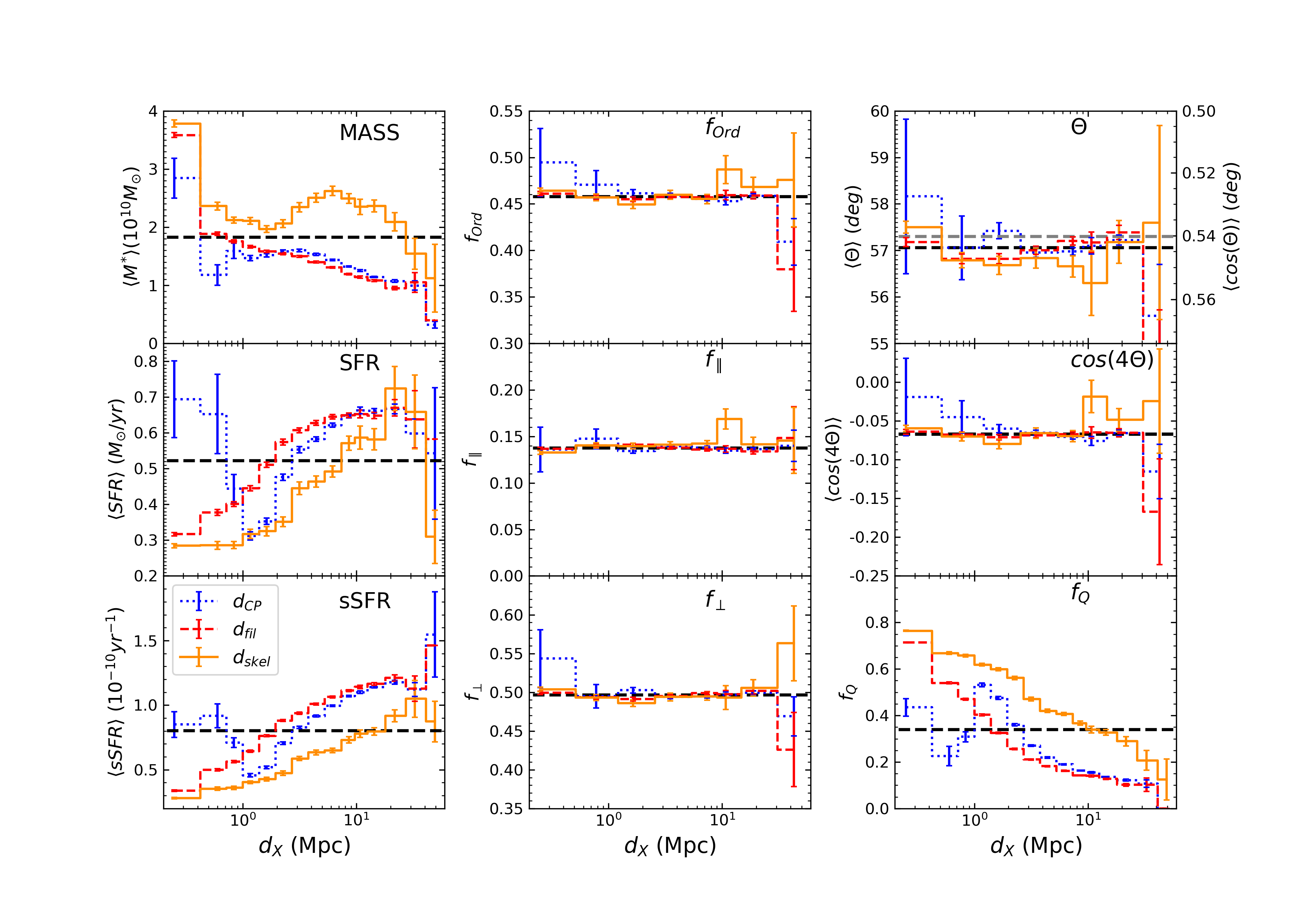}
\caption{Distributions of $\langle M^{\ast} \rangle$, $\langle \mathrm{SFR} \rangle$, $\langle \mathrm{sSFR} \rangle$, $\langle \theta \rangle$, $\langle \cos(4\theta) \rangle$, fraction of quenched galaxies ($f_{\mathrm{Q}}$), fraction of ordered galaxies ($f_{\mathrm{Ord}}$), fraction of parallel galaxies ($f_{\parallel}$), and fraction of perpendicular galaxies ($f_{\perp}$) as a function of the distances from the features of the cosmic web \dfil~(red dashed line in every panel), \dnode~(blue dotted line in every panel), and \dskel~(orange solid line in every panel). In the case of \dnode~ and \dskel, only galaxies outside of filaments (\dfil$\geq 1 \mathrm{Mpc}$) and inside filaments (\dfil$\leq 1 \mathrm{Mpc}$) have been considered, respectively. Error bars on the distributions have been computed through bootstrap resampling. The black dashed line in every panel is the average of the considered quantity in the full simulation box. In the top and middle panel of the right column, the grey line is $\bar{\theta}$ and $\overline{\cos(4\theta)}$, computed given equations \eqref{ptheta} and \eqref{pcos4t} for the total subhalo sample. In this figure, the grey dashed line corresponding to $\overline{\cos(4\theta)}$ is hidden behind the black dashed line in the same panel. Note that in every panel the first distance bin considered extends all the way to 0 for all distances, however it is cut due to the $x$-axis being in logarithmic scale. Note also that in the $f_{\parallel}$ and $f_{\perp}$ cases the y-axes of the plots cover very different ranges.}
\label{sfrmassangledist}
\end{figure*}

We derived the distributions of the average stellar mass ($M^{\ast}$), (specific-)SFR, $\theta$, $\cos(4\theta)$, and fractions of quenched, ordered, parallel, and perpendicular galaxies ($f_{\mathrm{Q}}, f_{\mathrm{Ord}}, f_{\parallel}, f_{\perp}$, respectively) as a function of $d_{\mathrm{fil}}$, $d_{\mathrm{CP}}$, and $d_{\mathrm{skel}}$ (Figure \ref{sfrmassangledist}). This figure offers a global and comprehensive view of how the various quantities vary with the considered distances. In the following, we describe the trends with distances from the features of the cosmic web for mass and SFR-related galaxy properties and spin-related galaxy properties separately.

\subsection{Mass and SFR-related quantities}
Stellar mass $\langle M^{\ast} \rangle$, $\langle \mathrm{SFR} \rangle$, $\langle \mathrm{sSFR} \rangle$, and $f_{\mathrm{Q}}$ all monotonically vary with \dfil, \dnode, and \dskel, with $\langle M^{\ast} \rangle$ and $f_{\mathrm{Q}}$ decreasing when moving away from structures and $\langle \mathrm{SFR} \rangle$ and $\langle \mathrm{sSFR} \rangle$ increasing. These trends reflect the fact that more massive galaxies inhabit the inner regions of structures, which are also the places where galaxy populations experience a larger degree of quenching. A decrease of the SFR with decreasing distance from the spine of filaments is also reported in observations \citep[e.g.][]{Bonjean2020, Kuutma2017}. However, $\langle M^{\ast} \rangle$ has a very similar trend with distances regardless of whether we consider a node or a filament (i.e. with respect to both \dnode~ and \dfil). Indeed the only difference between the curves is for trends with \dskel, i.e. galaxies inside filaments for which the mass is on average higher but varies less significantly when moving towards the nodes following the filamentary structures. The increase in the distribution of $\langle M^{\ast} \rangle$ at large \dskel~values is due to small number counts in the bins in this regime, due to our choice of a distance of \dfil$= 1$ Mpc to separate between the populations within and outside the core of the filaments (as \dskel is defined only for subhaloes with \dfil$\leq 1$ Mpc).  When a larger distance threshold is chosen, these bins become more populated and the distribution becomes monotonically decreasing and closer to the distributions for \dnode~ and \dfil. On the other hand, quantities related to SFR show some degree of difference between the structures, with the three curves for \dnode, \dfil, and \dskel~being separated. Therefore, structures seem to affect SFR differently than they do $M^{\ast}$, allowing for the former quantity to be used to better separate whether galaxies are close to filaments or to nodes. Also in the case of $\langle \mathrm{SFR} \rangle$, the increase seen at small \dnode~is due to these bins having low number counts due to our choice of a distance threshold to separate between galaxies in filaments and outside filaments (as \dnode~is defined only for subhaloes with \dfil$\geq 1$ Mpc). Choosing a smaller distance threshold has the effect of making these bins more populated and the $\langle \mathrm{SFR} \rangle$ distribution becomes monotonically increasing with \dnode.

\subsection{Spin-related quantities}

Spin-related quantities ($\langle \theta \rangle$, $\langle \cos(4\theta) \rangle$, $f_{\mathrm{Ord}}$, $f_{\parallel}$, and $f_{\perp}$) show little to no variation with respect to the distances from the various structures. In fact, often the various distributions are rather flat and overlapping. The only exception is the case of $\langle \theta \rangle$, which shows a flat distribution for \dfil, but not for \dskel. In this latter case, the value of $\langle \theta \rangle$ decreases moving away from nodes following the filaments. In the case of \dnode, there is a trend for a larger fraction of galaxies with their spin perpendicular to the filaments to be present (visible in an increased value for $\langle \theta \rangle$, $\langle \cos(4\theta) \rangle$, $f_{\mathrm{Ord}}$, and $f_{\perp}$) at small values of \dnode, but large error bars prevent us from confirming it. These results fit in the theoretical framework outlined by constrained tidal torque theory \citep{Codis2015}. As stated in the introduction, constrained tidal torque theory predicts the alignment of the spin of haloes with the filaments of the cosmic web. This alignment is subsequently changed to perpendicular by the further non-linear evolution of the filaments while they flow within the filaments. This can be due e.g. to the smooth accretion of vorticity-rich gas on to haloes that happen to be bigger than a given vorticity quadrant of the filaments (\citealt{Laigle2015vorticity}; see also \citealt{GaneshaiahVeenaCosmicBalletI}: accreting haloes are generally embedded in thinner filaments, and the subsequent accretion contributes to the spin becoming perpendicular to the filaments). As non-linear processes change halo spin while they flow within filaments towards clusters \citep{WangKang2017}, we expect a larger fraction of galaxies with spin parallel to the filaments far away from the clusters, which progressively reduces towards the clusters. In clusters we either expect a random orientation of spin (as these are multi-flow regions) or a preferentially perpendicular orientation of spin with respect to filaments (which our results seem to hint at).

\subsection{Galaxy properties variations with respect to a given cosmic web distance}

\begin{figure*}
\centering
\includegraphics[width = \linewidth]{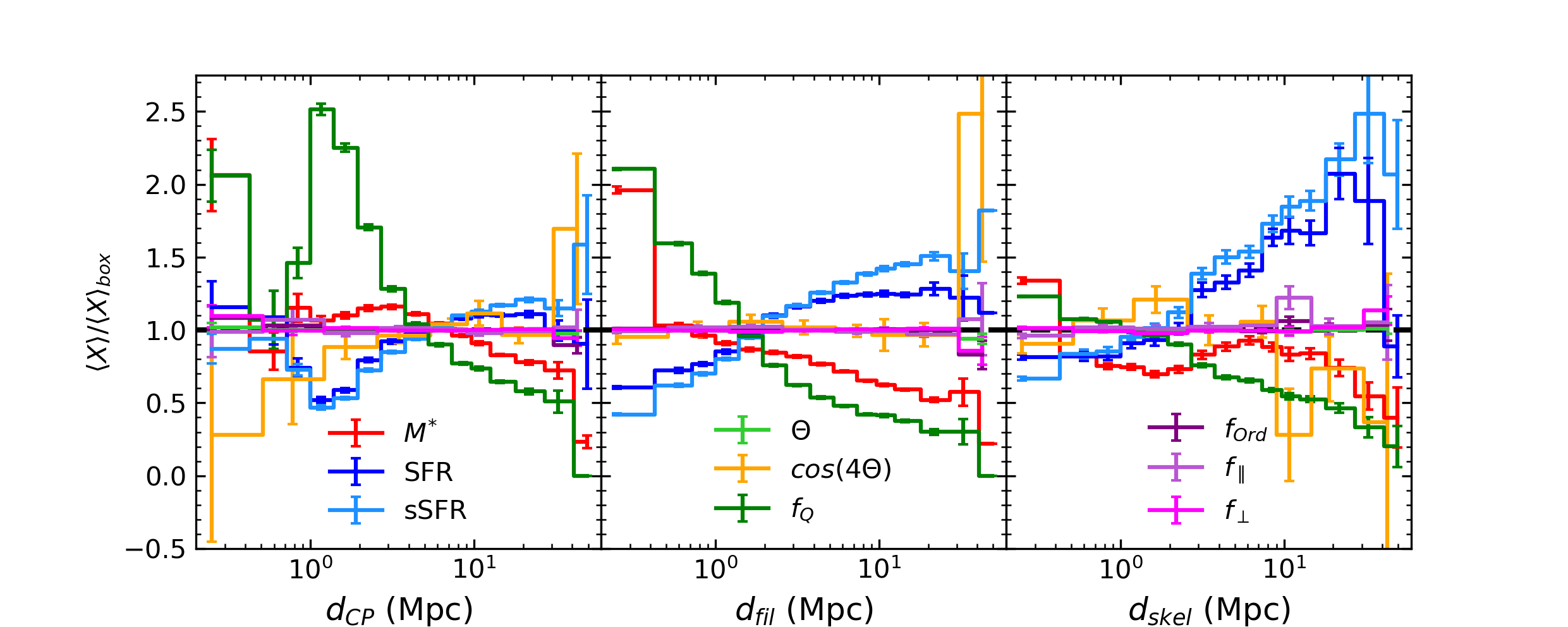}
\caption{Normalised distributions (expressed as $\frac{\langle X \rangle}{\langle X \rangle_{box}}$) with respect to distances \dnode~(left panel), \dfil~(middle panel), and \dskel~(right panel). The quantities considered in each panel ($\langle X \rangle$) are respectively: $\langle M^{\ast} \rangle$ (red), $\langle \mathrm{SFR} \rangle$ (dark blue), $\langle \mathrm{sSFR} \rangle$ (light blue), $\langle \theta \rangle$ (light green), $\langle \cos(4\theta) \rangle$ (orange), $f_{\mathrm{Q}}$ (dark green), $f_{\mathrm{Ord}}$ (dark purple), $f_{\parallel}$ (light purple), and $f_{\perp}$ (magenta). $\langle X \rangle_{box}$ indicates the average of the quantity taken including all subhaloes in the box (for \dfil), only those outside filaments (for \dnode), and only those inside filaments (for \dskel) over the full simulation volume. Error bars on the distributions have been computed through bootstrap resampling. Note that in every panel the first distance bin considered extends all the way to 0 for all distances, however it is cut due to the $x$-axis being in logarithmic scale.}
\label{distquantitiessamedist}
\end{figure*}

In order to compare how a given structure or infall path (nodes or filaments) is traced by each quantity, we normalised the distributions shown in Figure \ref{sfrmassangledist} to the average of each considered quantity, computed in the full extent of the box. In the case of average quantities like $\langle M^{\ast} \rangle$, $\langle \mathrm{SFR} \rangle$, $\langle \mathrm{sSFR} \rangle$, $\langle \theta \rangle$, and $\langle \cos(4\theta) \rangle$, we divided the distributions shown in Figure \ref{sfrmassangledist} by the average of the same quantities computed using all the subhaloes in the box. In the case of fraction quantities, i.e. $f_{\mathrm{Q}}$, $f_{\mathrm{Ord}}$, $f_{\parallel}$, and $f_{\perp}$, we divided the distributions of Figure \ref{sfrmassangledist} by the fractions of the same quantities computed using all subhaloes in the box. The trends of the quantities shown in the panels of Figure \ref{sfrmassangledist} for \dfil, \dnode, and \dskel~are shown in Figure \ref{distquantitiessamedist}.

The normalised distributions show how different galaxy properties trace the same type of structure. In this regard, when \dnode~is considered, $f_{\mathrm{Q}}$ seems to be the quantity which shows the greatest variation across the range of considered distances. This remains generally true also for \dfil, although in this case also the SFR and sSFR acquire importance, especially to trace large distances from filaments. Finally, these two latter quantities are those that better trace \dskel~among all. In the context of spin-related quantities, only $\cos(4\theta)$ shows a variation comparable to other quantities, at small \dnode, although with large error bars.

\subsection{Global vs. local environment}
We explore the relation between the global environment of galaxies (i.e. their position with respect to LSS features) and their local environment (i.e. the environmental density in their immediate vicinity, regardless of the type of structures in which they are embedded). In our case, the global environment of galaxies is explored through the already-introduced distances (\dnode, \dfil, and \dskel). The local environment is instead codified through the DTFE density value at the position of each galaxy (in $\mathrm{Mpc}^{-3}$), directly measured by \disperse~ and used in the derivation of the skeleton. 

\begin{figure}
\centering
\includegraphics[width = \linewidth]{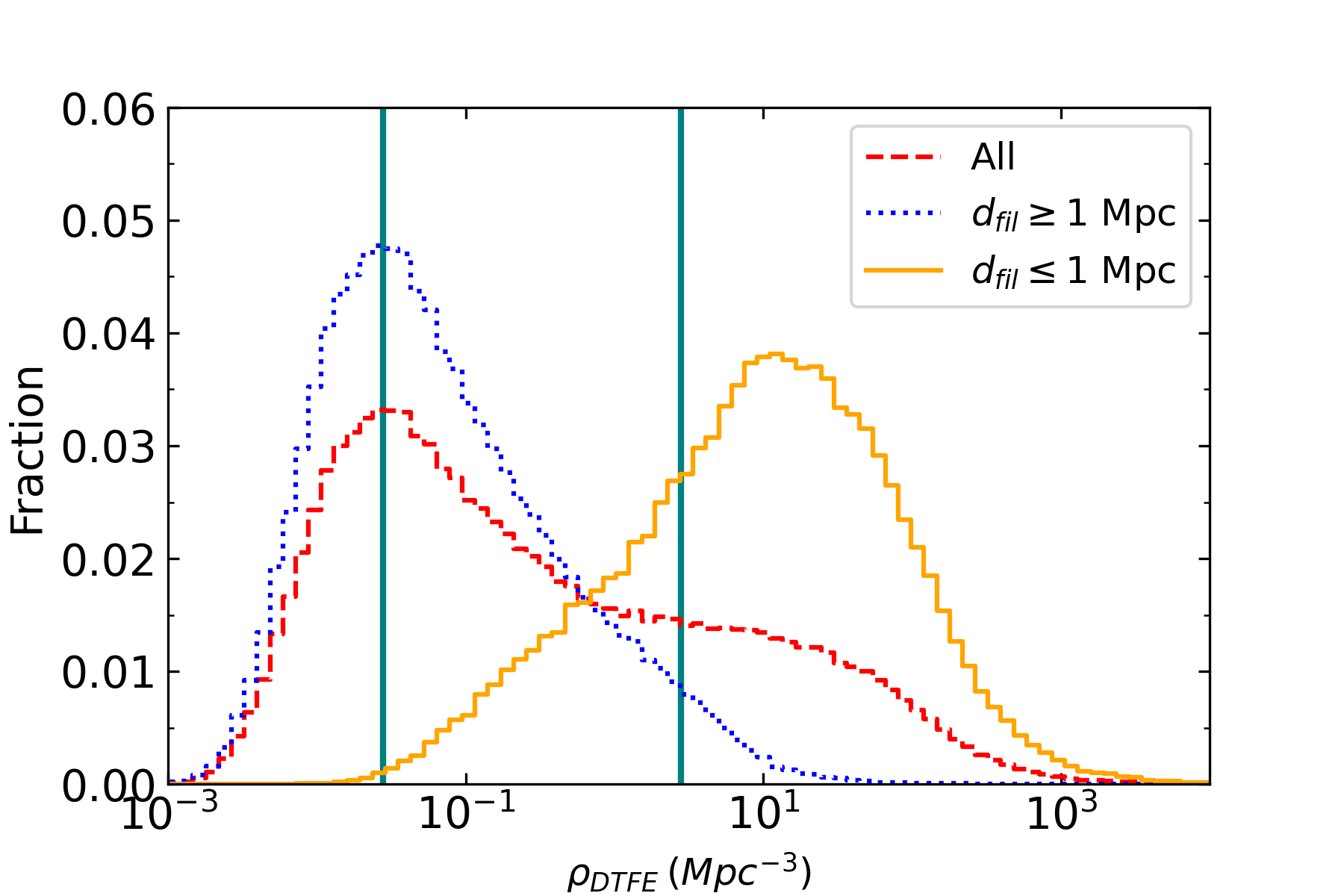}
\caption{Distributions of density ($\rho_{\mathrm{DTFE}}$) as measured through the DTFE value output by \disperse~at the position of each galaxy. The red dashed line refers to the total population of galaxies, the blue dotted line to the galaxies outside of filaments and the yellow solid line to the galaxies inside filaments. The two vertical blue lines represent the 25th and 75th percentiles of the total distribution.}
\label{dtfe_density_dist}
\end{figure}

Figure \ref{dtfe_density_dist} shows the distributions of density values for all galaxies and for those inside and outside of filaments. The distribution for all galaxies shows a peak at small density values (slightly above $10^{-2} \mathrm{Mpc}^{-3}$) and decreases slowly, reaching large values of $\sim 10^{3} \mathrm{Mpc}^{-3}$. Selecting only galaxies inside or outside filaments has the effect of restricting the density range considered (the distribution for galaxies inside filaments indeed peaks at higher values $\gtrsim 10^{1} \mathrm{Mpc}^{-3}$, while the distribution for galaxies outside filaments peaks at $\gtrsim 10^{-2} \mathrm{Mpc}^{-3}$). This confirms that galaxies outside filaments are in environments which are on average locally less dense than those of galaxies within filaments. However, the large overlap between the local density distributions of galaxies inside and outside filaments strongly supports the fact that local density alone is not a good criterion to separate the features of the cosmic web and that a topological definition has to be used to detect the filaments. This overlap in local density for galaxies belonging to different cosmic structures opens the question of whether the trends that we detect and we show above are due to local density or to the anisotropic properties of the filaments. Indeed, if galaxies inside and outside filaments can experience the same kind of local environment while belonging to different types of global LSS features (and similarly for galaxies inside clusters or on their outskirts), then it could mean that galaxies outside and inside cosmic web structures can share the same properties if these are driven predominantly by local density. 

\begin{figure}
\centering
\includegraphics[width = \linewidth]{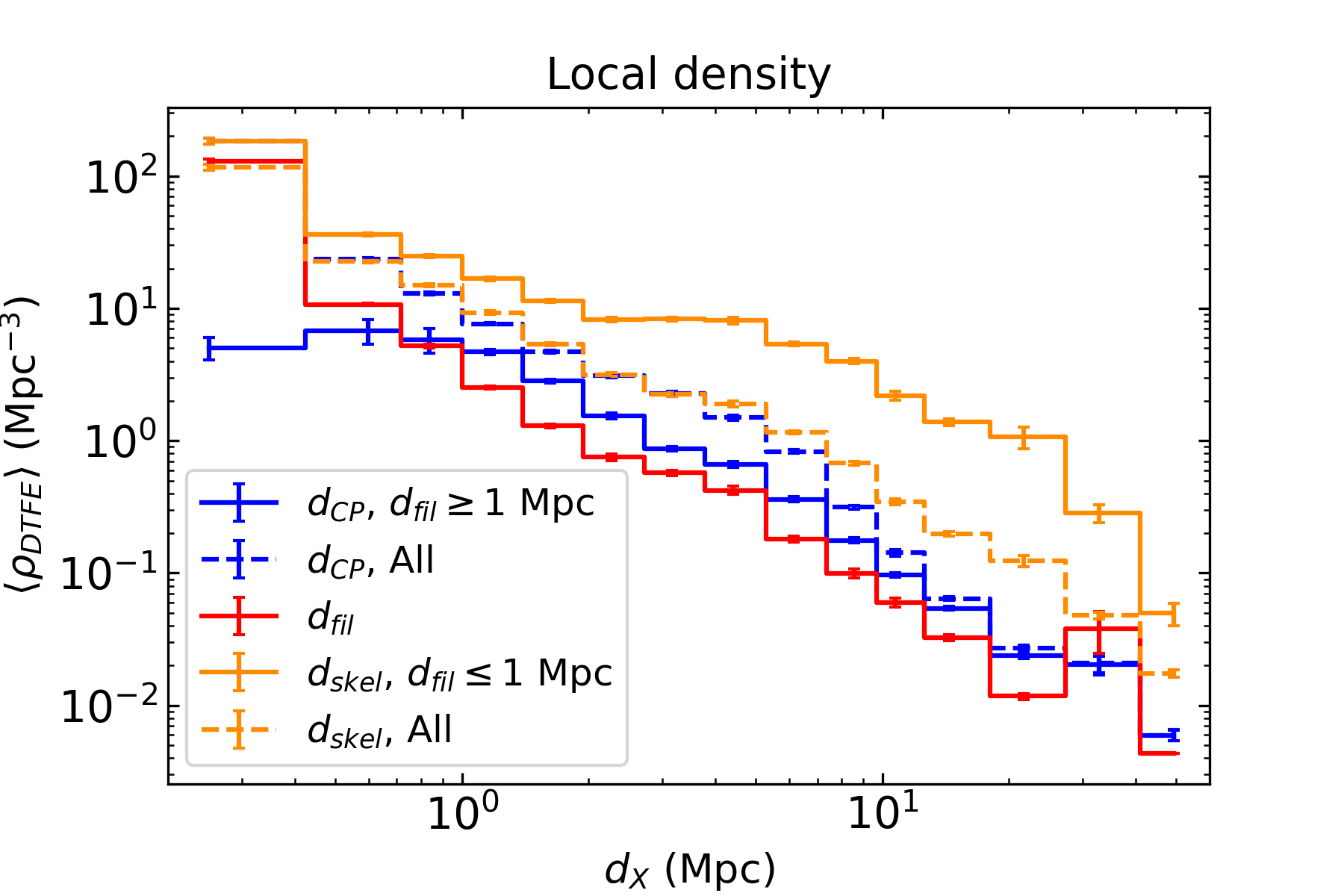}
\caption{Distributions of $\langle \rho_{\mathrm{DTFE}} \rangle$ as a function of the distances from the features of the cosmic web \dfil~ (red line), \dnode~ (blue line), and \dskel~ (orange line). In the case of \dnode~ and \dskel, solid lines refer to galaxies outside of filaments (\dfil$\geq 1 \mathrm{Mpc}$) and inside filaments (\dfil$\leq 1 \mathrm{Mpc}$), respectively, while dashed lines refer to the total galaxy population. Error bars on the distributions have been computed through bootstrap resampling. Note that in every panel the first distance bin considered extends all the way to 0 for all distances, however it is cut due to the $x$-axis being in logarithmic scale.}
\label{density_dist_func}
\end{figure}

Figure \ref{density_dist_func} shows how the average local density $\langle \rho_{\mathrm{DTFE}} \rangle$ varies as a function of the considered distances from  the structures. As expected, local density decreases monotonically, moving away from structures. For galaxies within filaments, density is consistently higher than in other cases, and seems to present a milder decrease as well. As different cosmic web features broadly cover different density ranges, any dependence of galaxy properties on local density rather than on their global environment could mimic a trend with the distances from the cosmic web features as those shown above. In order to check the effect of local density on our conclusions we perform the following analysis.

If we derive the same distributions as in Figure \ref{sfrmassangledist} separated between high- and low-density galaxies (not shown here), we find that in general galaxies follow the expected trends in the high- and low-density case, although they are less defined than when the global galaxy population is considered. In general, high-density galaxies are less star-forming and show a higher fraction of quenched systems. Both high- and low-density galaxies show trends with the various distances from structures, with the (specific-)SFR increasing and the fraction of quenched galaxies decreasing with increasing \dnode, \dfil, and \dskel. Spin-related quantities do now show any difference between high- and low-density galaxies and no trends with the distances from structures seem to be visible when galaxies are separated according to their local environment. We separate local high- and low-density environments using the 75th and the 25th percentiles of the local density distribution, respectively (shown in Figure \ref{dtfe_density_dist}).

\begin{figure*}
\centering
\includegraphics[trim = 2cm 1cm 1cm 1cm, clip = true, width = \linewidth]{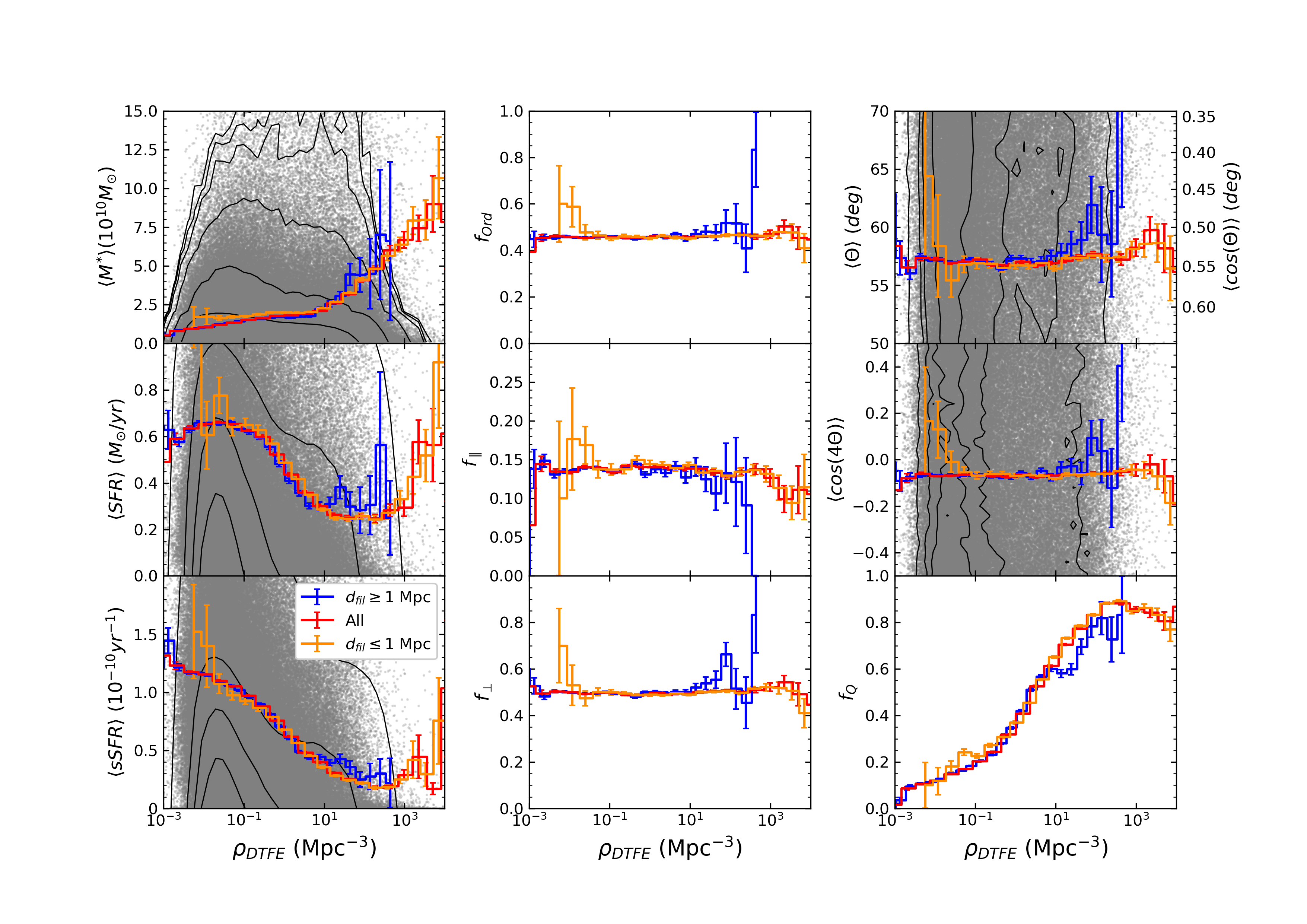}
\caption{Distributions of $\langle M^{\ast} \rangle$, $\langle \mathrm{SFR} \rangle$, $\langle \mathrm{sSFR} \rangle$, $\langle \theta \rangle$, $\langle \cos(4\theta) \rangle$, fraction of quenched galaxies ($f_{\mathrm{Q}}$), fraction of ordered galaxies ($f_{\mathrm{Ord}}$), fraction of parallel galaxies ($f_{\parallel}$), and fraction of perpendicular galaxies ($f_{\perp}$) as a function local density ($\rho_{\mathrm{DTFE}}$) for all galaxies (red line in every panel), galaxies inside filaments (\dfil$\leq 1 \mathrm{Mpc}$, orange line in every panel), and outside of filaments (\dfil$\geq 1 \mathrm{Mpc}$, blue line in every panel). Error bars on the distributions have been computed through bootstrap resampling. Where present, gray points represent the scatter of the total galaxy population in  the considered quantity-local density plane (black contours show the shape of the distribution where points saturate). Note that in the $f_{\parallel}$ and $f_{\perp}$ cases the y-axes of the plots cover very different ranges.}
\label{distributions_densfunc}
\end{figure*}

Figure \ref{distributions_densfunc} shows the dependence of galaxy properties on the local density. As expected, trends with local density are visible for mass and SFR-related quantities. As density increases, galaxy $M^{\ast}$ also increases, while SFR and sSFR decrease. Correspondingly, in denser local environments the fraction of quenched galaxies is higher. On the other hand, no trend with local density can be seen for spin-related quantities. Moreover, the distributions for galaxies inside filaments and outside filaments show no difference, regardless of the quantity considered.

In order to disentangle the effect of local density from the effect of structures we perform the following test: we divide the galaxy sample in local density bins and we shuffle the considered galaxy quantity (mass, SFR, etc) 1000 times within the density bin while keeping the distances to the LSS features (\dnode, \dfil, \dskel) unchanged. This eliminates the relation between a given quantity and the distances to structures while keeping intact the relation between the given quantity and local density (thanks to the mixing of the galaxy property values among galaxies within the same local density bin). In this way, if galaxies in a given local density bin have different properties than elsewhere (e.g. lower SFR on average), this will be preserved, while the shuffling eliminates the relation between the given quantity and the distances to the cosmic web features.

The distributions of mass and SFR-related quantities do not change visibly when galaxies are re-shuffled. In these cases we cannot exclude that the trends that we see are due to how local density varies in response to the distance to the various structures, rather than to the structures themselves. On the other hand, for spin-related quantities and in particular the amount of galaxies with their spin perpendicular to the filaments, we see variations in the distributions of re-shuffled galaxies, in particular at small distances from nodes for galaxies outside filaments. It is an indication that the spin of galaxies could be a way to trace the cosmic web which is more independent of local density than other quantities.

\begin{figure*}
\centering
\includegraphics[trim = 1cm 1cm 2cm 0cm, clip = true, width = \linewidth]{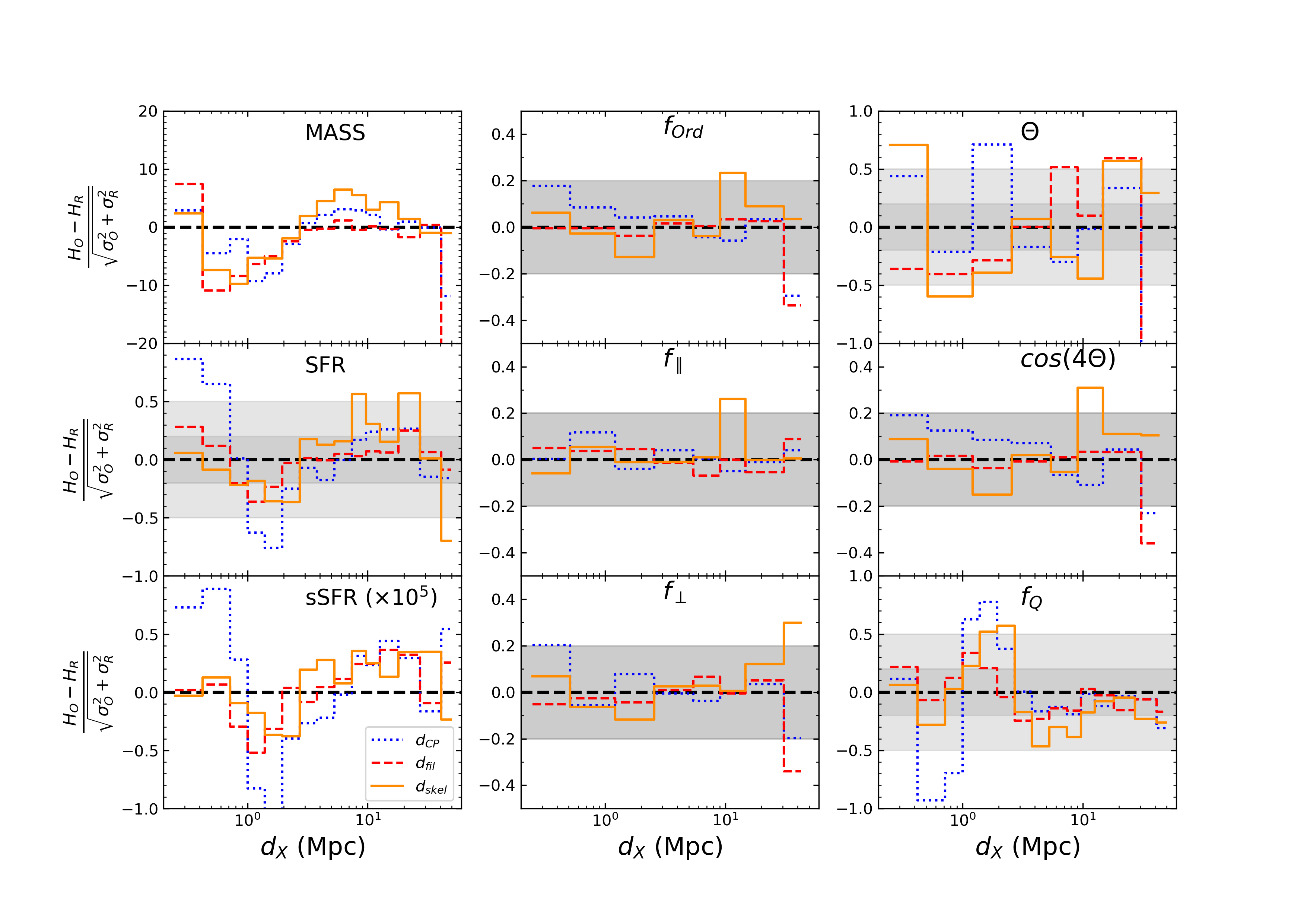}
\caption{Differences between the original distributions $H_{O}$ and the re-shuffled ones $H_{R}$, normalised to the sum in quadrature of their errors ($\sigma_{O}$ and $\sigma_{R}$, respectively) for the quantities $\langle M^{\ast} \rangle$, $\langle \mathrm{SFR} \rangle$, $\langle \mathrm{sSFR} \rangle$, $\langle \theta \rangle$, $\langle \cos(4\theta) \rangle$, fraction of quenched galaxies ($f_{\mathrm{Q}}$), fraction of ordered galaxies ($f_{\mathrm{Ord}}$), fraction of parallel galaxies ($f_{\parallel}$), and fraction of perpendicular galaxies ($f_{\perp}$) as a function of the distances from the features of the cosmic web \dfil~ (red dashed line in every panel), \dnode~ (blue dotted line in every panel), and \dskel~ (orange solid line in every panel). In the case of \dnode~ and \dskel, only galaxies outside of filaments (\dfil$\geq 1 \mathrm{Mpc}$) and inside filaments (\dfil$\leq 1 \mathrm{Mpc}$) have been considered, respectively. In all panels, except for $\langle M^{\ast} \rangle$ and $\langle \mathrm{sSFR} \rangle$ (where they would have encompassed the whole range on the $y$-axis), light grey and dark grey areas show the $0.5\sigma$ and $0.2\sigma$ range, respectively.}
\label{ratio_reshuffled}
\end{figure*}

We have tried to quantify the difference between the re-shuffled and the original distributions by means of the quantity $\frac{H_{O}-H_{R}}{\sqrt{\sigma_{O}^{2}+\sigma_{R}^{2}}}$, where $H_{O}$ are the original distributions shown in Figure \ref{sfrmassangledist}, $H_{R}$ the re-shuffled ones, and $\sigma_{O}$ and $\sigma_{R}$ their uncertainties. This quantity highlights any significant change between the original and re-shuffled distributions and it is shown in Figure \ref{ratio_reshuffled}. We find that in the case of $\langle \theta \rangle$, differences among the distributions are as high as $0.5\sigma$, while in the case of $\langle \cos(4\theta) \rangle$, $f_{\mathrm{Ord}}$, $f_{\parallel}$, and $f_{\perp}$ they are in the range of $0.2\sigma$. Although the differences between the original and re-shuffled distributions of spin-related quantities are not large, still they are detected. In particular, in the case of $\langle \cos(4\theta) \rangle$ and $f_{\mathrm{Ord}}$ when considered with respect to \dnode, the differences are consistently positive, meaning that the re-shuffled distributions have a smaller amplitude than the original ones. This is in sharp contrast with the case of $\langle \mathrm{sSFR} \rangle$, where the differences between the re-shuffled and the original distributions are as low as $10^{-5}\sigma$. This supports the hypothesis that, although trends of spin-related quantities with distances from the cosmic web are more difficult to detect, they may be more insensitive to the local density of galaxies.

In the case of $\langle \mathrm{SFR} \rangle$ and $f_{\mathrm{Q}}$, the differences between the distributions are also in the range of $0.5\sigma$ as for $\langle \theta \rangle$, however they oscillate around zero, without being consistently positive or negative across the range of distances considered. The largest differences between re-shuffled and original distributions are observed in the case of mass (of the order of $10\sigma$) however this could be due to the precision with which stellar mass is measured.

\subsection{Mass dependence}
\label{results_refined}
\begin{figure*}
\centering
\includegraphics[trim = 2cm 1cm 1cm 0cm, clip = true, width = \linewidth]{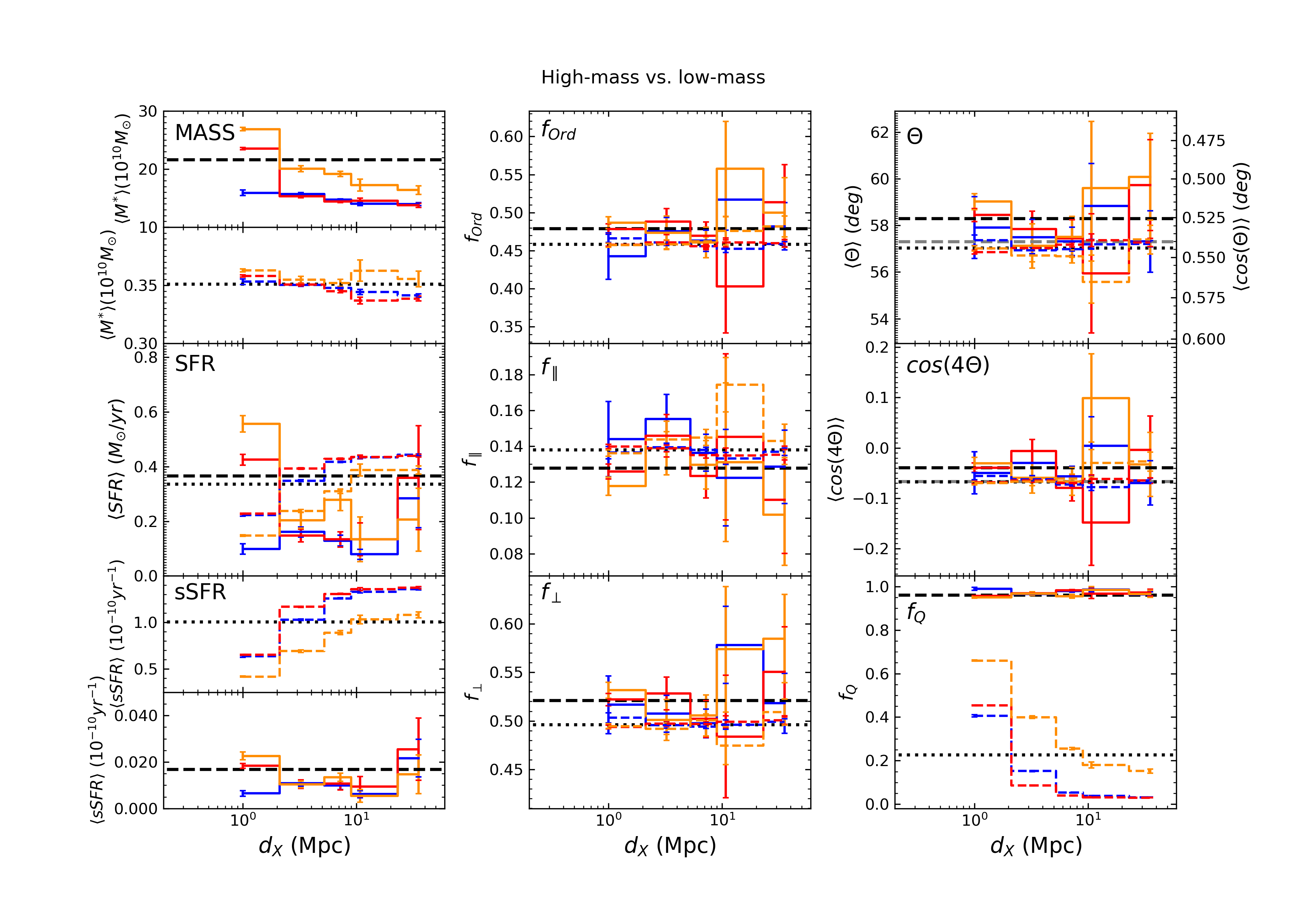}
\caption{Distributions of $\langle M^{\ast} \rangle$, $\langle \mathrm{SFR} \rangle$, $\langle \mathrm{sSFR} \rangle$, $\langle \theta \rangle$, $\langle \cos(4\theta) \rangle$, fraction of quenched galaxies ($f_{\mathrm{Q}}$), fraction of ordered galaxies ($f_{\mathrm{Ord}}$), fraction of parallel galaxies ($f_{\parallel}$), and fraction of perpendicular galaxies ($f_{\perp}$) as a function of the distances from the features of the cosmic web \dfil~ (red line in every panel), \dnode~ (blue line in every panel), and \dskel~ (orange line in every panel) and split between high- and low-mass galaxies. In the case of \dnode~ and \dskel, only galaxies outside of filaments (\dfil$\geq 1 \mathrm{Mpc}$) and inside filaments (\dfil$\leq 1 \mathrm{Mpc}$) have been considered, respectively. Error bars on the distributions have been computed through bootstrap resampling. The black dashed (dotted) line in every panel is the average of the considered quantity in the full simulation box considering only high-mass (low-mass) galaxies. In the top and middle panel of the central column, the grey line is $\bar{\theta}$ and $\overline{\cos(4\theta)}$, computed given equations \eqref{ptheta} and \eqref{pcos4t} for the total subhalo sample. In each panel, solid lines refer to high-mass galaxies and dashed lines to low-mass galaxies. In the case of mass and sSFR, the panel has been split in two to take into account the very different ranges on the $y$-axis occupied by the distributions. Note that in every panel the first distance bin considered extends all the way to 0 for all distances, however it is cut due to the $x$-axis being in logarithmic scale.}
\label{sfrmassangledistmass}
\end{figure*}

We check how the discovered trends depend on galaxy stellar mass. Indeed stellar mass has an influence on the SFR of galaxies, with more massive galaxies also being the less star-forming, possibly due to their having been formed earlier (an effect known as ``downsizing'', see e.g. \citealt{Cowie1996, Bolzonella2010, Renzini2006, Thomas2005, Pozzetti2010, Cimatti2006, Cucciati2006}). Moreover, spin alignment also has a dependance on mass, with more massive galaxies being perpendicular to filaments and lower mass galaxies being preferentially aligned, as we outlined in section \ref{galpropsspin}. In order to check whether the trends we observe are due specifically to galaxies in a given mass range (either high-mass or low-mass), we derive the distributions shown in Figure \ref{sfrmassangledist} for high- and low-mass galaxies separately. This is shown in Figure \ref{sfrmassangledistmass}.

When galaxies are divided according to mass, the most striking feature is that in the case of SFR-related quantities (e.g. SFR, sSFR, and $f_{\mathrm{Q}}$) a trend with the distances from the structures is visible only for low-mass galaxies. High-mass galaxies generally show a lower amount of (specific-)SFR and a higher quenched fraction, but the distributions are uniform throughout the ranges of distances explored. On the other hand, for low-mass galaxies the same trends with distances as for the general population are recovered. This is in agreement with a scenario in which quenching processes are separable between mass- and environment-driven \citep[e.g.][]{Peng2010}, with environmental quenching affecting primarily low-mass galaxies and mass quenching affecting primarily high-mass galaxies. Indeed, for low-mass galaxies, star-formation activity decreases with decreasing distance from the nodes (either following the filaments or considered isotropically) and decreasing distance from the axis of the filaments. However, the possibility to distinguish between structures is reduced, with the distributions with respect to \dfil~ and \dnode~ largely overlapping. In the case of mass distributions, weak trends with distances from structures are maintained both for high- and low-mass galaxies, with the curves largely overlapping. The only exceptions are the $\langle M^{\ast} \rangle$ distribution for high-mass galaxies and the $\langle \mathrm{SFR} \rangle$, $\langle \mathrm{sSFR} \rangle$, and $f_{\mathrm{Q}}$ distributions for low-mass galaxies as a function of \dskel, which show how massive galaxies in filaments have an enhanced mass and a low-mass galaxies have a lower amount of SFR with respect to the remaining galaxy population. Spin-related quantities generally do not show differences between high- and low-mass galaxies, except for hints of the distributions of high-mass galaxies being shifted at larger values of $\theta$, with a possible faint trend for $\theta$ to decrease with \dfil~ and \dskel. This is consistent with the spin of high-mass galaxies being more perpendicular to filaments in general and especially close to structures and also fits within the theoretical framework laid out in \citet{Laigle2015vorticity}, where gas accretion onto haloes bigger than the vorticity quadrant of filaments can affect the direction of the spin. In our case, galaxies are more likely to cross several vorticity quadrants at small \dfil, while massive galaxies have had more time to accrete matter in this environment.

\begin{figure*}
\centering
\includegraphics[width = \linewidth]{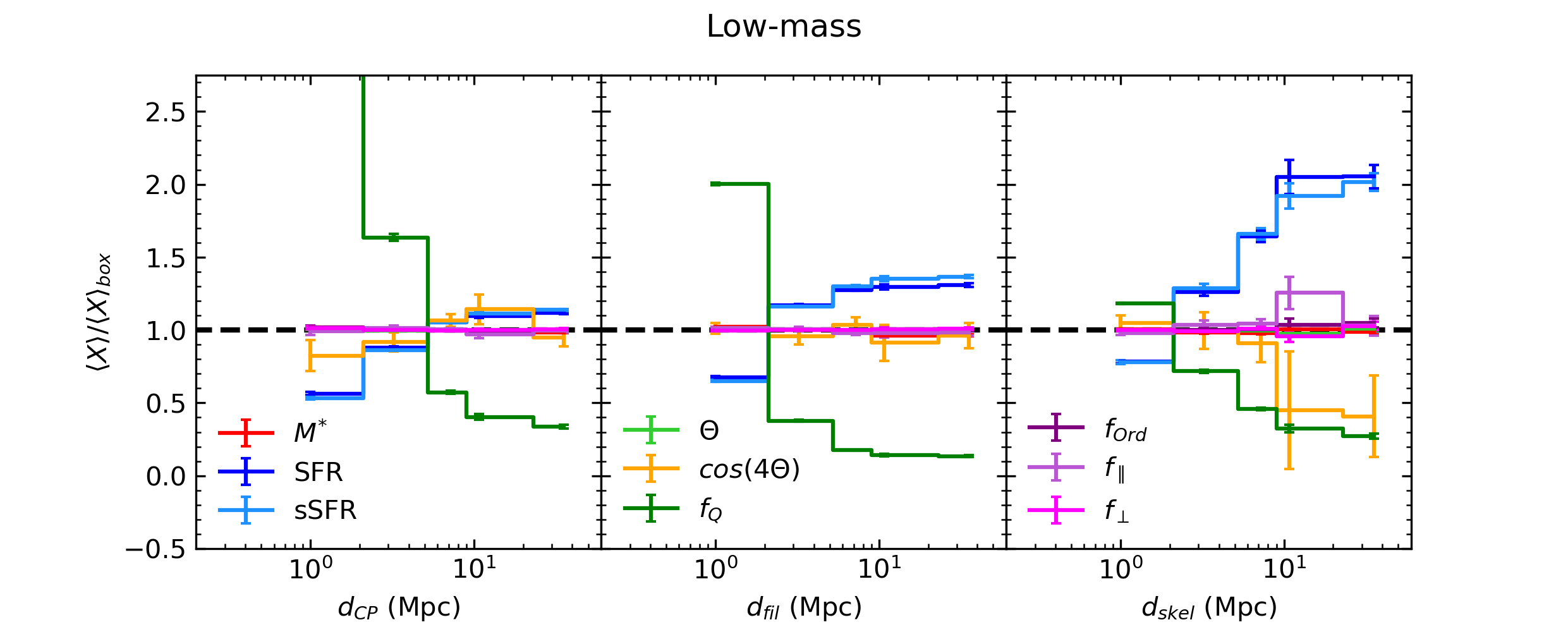}
\caption{Normalised distributions (expressed as $\frac{\langle X \rangle}{\langle X \rangle_{box}}$) with respect to distances \dnode~ (left panel), \dfil~ (middle panel), and \dskel~ (right panel). The quantities considered in each panel ($\langle X \rangle$) are respectively: $\langle M^{\ast} \rangle$ (red), $\langle \mathrm{SFR} \rangle$ (dark blue), $\langle \mathrm{sSFR} \rangle$ (light blue), $\langle \theta \rangle$ (light green), $\langle \cos(4\theta) \rangle$ (orange), $f_{\mathrm{Q}}$ (dark green), $f_{\mathrm{Ord}}$ (dark purple), $f_{\parallel}$ (light purple), and $f_{\perp}$ (magenta). $\langle X \rangle_{box}$ indicates the average of the quantity taken including all low-mass subhaloes in the box (for \dfil), only those outside filaments (\dnode), and only those inside filaments (\dskel) over the full simulation volume. This figure only refers to low-mass galaxies. Error bars on the distributions have been computed through bootstrap resampling. Note that in every panel the first distance bin considered extends all the way to 0 for all distances, however it is cut due to the $x$-axis being in logarithmic scale.}
\label{distquantitiessamedistmass}
\end{figure*}

In Figure \ref{distquantitiessamedistmass} we show the normalised distributions of galaxy properties as a function of the distance to the structures. As several trends that were visible for the general population are absent in the case of high-mass galaxies, we only show the distributions for low-mass galaxies. In this case, trends with the distances from the LSS features are preserved and SFR-related quantities are those that show the largest variation with the distances with respect to structures, in agreement with what previously found.

\subsection{Dividing galaxies by spin parameter}
We further refine our analysis by looking at the two distinct populations of high-spin parameter and low-spin parameter galaxies, divided using the percentiles of the Bullock parameter distribution and introduced in Section \ref{galpropsspin}. Given that Figure \ref{thetamassdist} showed how high-spin parameter galaxies are those which carry most of the signal related to spin alignment, especially when high- and low-mass galaxies are considered separately, in the following we focus only on this sub-sample.

In this case, the situation is not dissimilar than what obtained in Figure \ref{sfrmassangledistmass} for the total galaxy population. Trends with the distances from structures are still visible for $M^{\ast}$ and SFR-related quantities, but the large error bars and a large overlapping of the distributions of spin-related quantities prevent us from detecting any secure trend.

\begin{figure*}
\centering
\includegraphics[width = \linewidth]{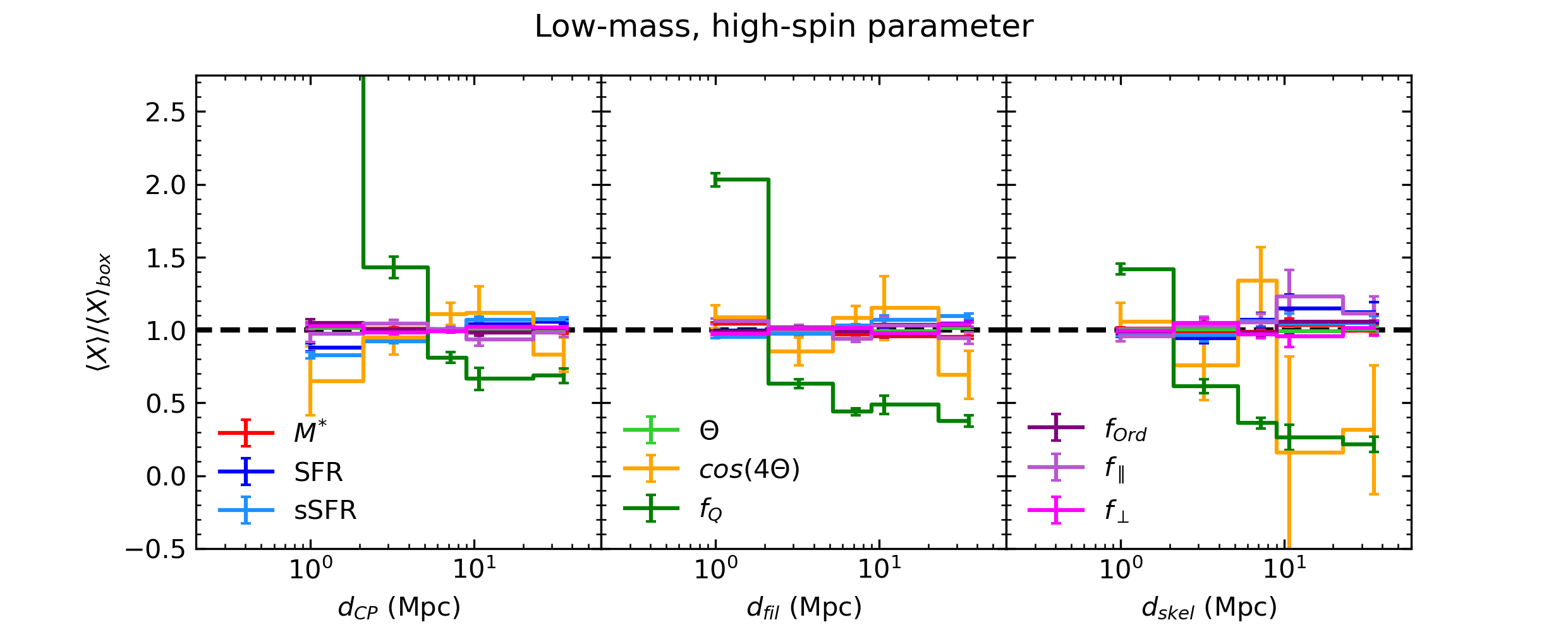}
\caption{Normalised distributions (expressed as $\frac{\langle X \rangle}{\langle X \rangle_{box}}$) with respect to distances \dnode~ (left panel), \dfil~ (middle panel), and \dskel~ (right panel). The quantities considered in each panel ($\langle X \rangle$) are respectively: $\langle M^{\ast} \rangle$ (red), $\langle \mathrm{SFR} \rangle$ (dark blue), $\langle \mathrm{sSFR} \rangle$ (light blue), $\langle \theta \rangle$ (light green), $\langle \cos(4\theta) \rangle$ (orange), $f_{\mathrm{Q}}$ (dark green), $f_{\mathrm{Ord}}$ (dark purple), $f_{\parallel}$ (light purple), and $f_{\perp}$ (magenta). $\langle X \rangle_{box}$ indicates the average of the quantity taken including all low-mass, high-spin parameter subhaloes in the box (for \dfil), only those outside filaments (\dnode), and only those inside filaments (\dskel) over the full simulation volume. Only low-mass, high-spin parameter galaxies are considered in this figure. Note that in every panel the first distance bin considered extends all the way to 0 for all distances, however it is cut due to the $x$-axis being in logarithmic scale.}
\label{distquantitiessamedistmassfastrot}
\end{figure*}

Figure \ref{distquantitiessamedistmassfastrot} shows the normalised distributions of galaxy properties for low-mass galaxies as in Figure \ref{distquantitiessamedistmass}, for the case when only high-spin parameter galaxies are considered. Trends with the distances from the structures are again visible for low-mass, high-spin parameter galaxies. The SFR-related quantity $f_{\mathrm{Q}}$ is indeed the quantity that shows the largest variation with respect to the distances from the structures, matched, in the case of \dskel~ and \dfil, also by $\langle \cos(4\theta) \rangle$.

\section{Concluding remarks}
\label{conclusions}
In this work we use the IllustrisTNG simulation, coupled with the \disperse~cosmic web extractor, to analyse how different features of the cosmic web (clusters and filaments) affect galaxy properties (mass, star-formation, and the direction of the angular momentum vector). Although the relation of these properties to the features of the cosmic web has been analysed in great detail in the literature independently for the three quantities, it is the first time that they are compared and contrasted in an extensive and comprehensive way. In particular we explore the possibility that one or more galaxy properties show different trends with the various cosmic web features and present a science case for their use to improve the detection of the cosmic web. Based on our analysis, we are able to draw the following conclusions:

\begin{enumerate}
\item When the distributions of galaxy properties are considered separately, SFR-related quantities allow to distinguish between \dfil~ (proxy for the accretion onto filaments) and \dnode~ and \dskel, proxies for the accretion onto nodes (further separated between the isotropic case and flowing inside the filaments). Mass and spin-related quantities seem to allow only for a distinction between \dnode~ and \dfil~ with respect to \dskel.

\item When the three distances are considered separately, the distributions of SFR-related quantites are those that show the largest variation with respect to each distance, a further confirmation that they are the best tracers for \dnode, \dfil, and \dskel. Mass shows a smaller variation, confirming as the second choice for a tracer quantity, and spin-related quantities show the lowest amount of variation and are therefore the worse tracer.

\item SFR-related quantities show also a large dependence on the local environment of galaxies, which may prevent their use as tracers to improve the detection of the cosmic web. On the other hand, spin related quantities are more robust with respect to the effect of local density. Although the strength of the signal of the recovered trends is lower, their use could provide a detection of the cosmic web in a way more independent from the local density.

\item When galaxies are separated by mass, the trends explored in the general case are visible mainly for the low-mass galaxy population. However, also in this case, SFR-related quantities are those that allow for the best separation between the distances to the various structures.

\item If only high-spin parameter subhaloes are selected and further separated in mass, trends in the distribution of quantities with the considered distances are more difficult to detect. Indeed, they seem to be visible only in the case of \dnode~ and \dskel, where $f_{\mathrm{Q}}$ and $\cos(4\theta)$ trace structures equally well.

\end{enumerate}

With a renewed interest in the study of the cosmic web and several large-scale upcoming galaxy surveys underway, it is vital to improve our understanding of both how the cosmic web affects galaxy evolution and how we can better use galaxy properties to detect the cosmic web. The results exposed here will greatly benefit from larger samples of subhaloes obtained from future, larger simulative efforts, which will allow to increase the statistical significance of the trends which are most difficult to detect (e.g. those related to the direction of the angular momentum of galaxies). In addition, further analyses targeting these same trends for particular sub-samples of simulated galaxies (e.g. in certain magnitude ranges or separated by morphological type) which reproduce observed data sets, will ensure the possibility to apply this context to future expected surveys. We also aim to perform similar analyses to what presented in this work while also better characterising the various structures of the cosmic web, e.g. differentiating between thick and thin filaments. Our final goal is to apply this framework to real data sets in the future.

\begin{acknowledgements}
We would like to thank the anonymous referee for the careful comments, which improved the quality of the paper.

NM would like to thank Louis Legrand, Tony Bonnaire, and Alexander Kolodzig for useful and fruitful discussions during the performing of the analysis exposed in this work. 

This research has been supported by the funding for the ByoPiC project from the European Research Council (ERC) under the European Union's Horizon 2020 research and innovation programme grant agreement ERC-2015-AdG 695561.

We would like to thank the IllustrisTNG team for publicly releasing the full simulated snapshots and halo catalogues.

NM would like to thank Thierry Sousbie and Christophe Pichon for developing and making freely available the \disperse~code.

\end{acknowledgements}

\bibliographystyle{aa}
\bibliography{spin}

\appendix

\section{Thresholding the critical points density}
\label{appendix_thresh_cpdens}
In our analysis we have made use of the critical points identified by \disperse~(bifurcations and maxima) as a proxy for galaxy overdensities (groups and clusters). However, the identification and matching of these points with confirmed galaxy clusters is not easy, as the distribution of these points depends to some extent on the smoothing and persistence threshold adopted \citep[see e.g.][]{Malavasi2020b}. For this reason, considering all maxima and bifurcations as potential clusters or groups may be misleading. In order to check how much this affects our analysis, we have thresholded the critical points according to their density.

\begin{figure}
\centering
\includegraphics[width = \linewidth]{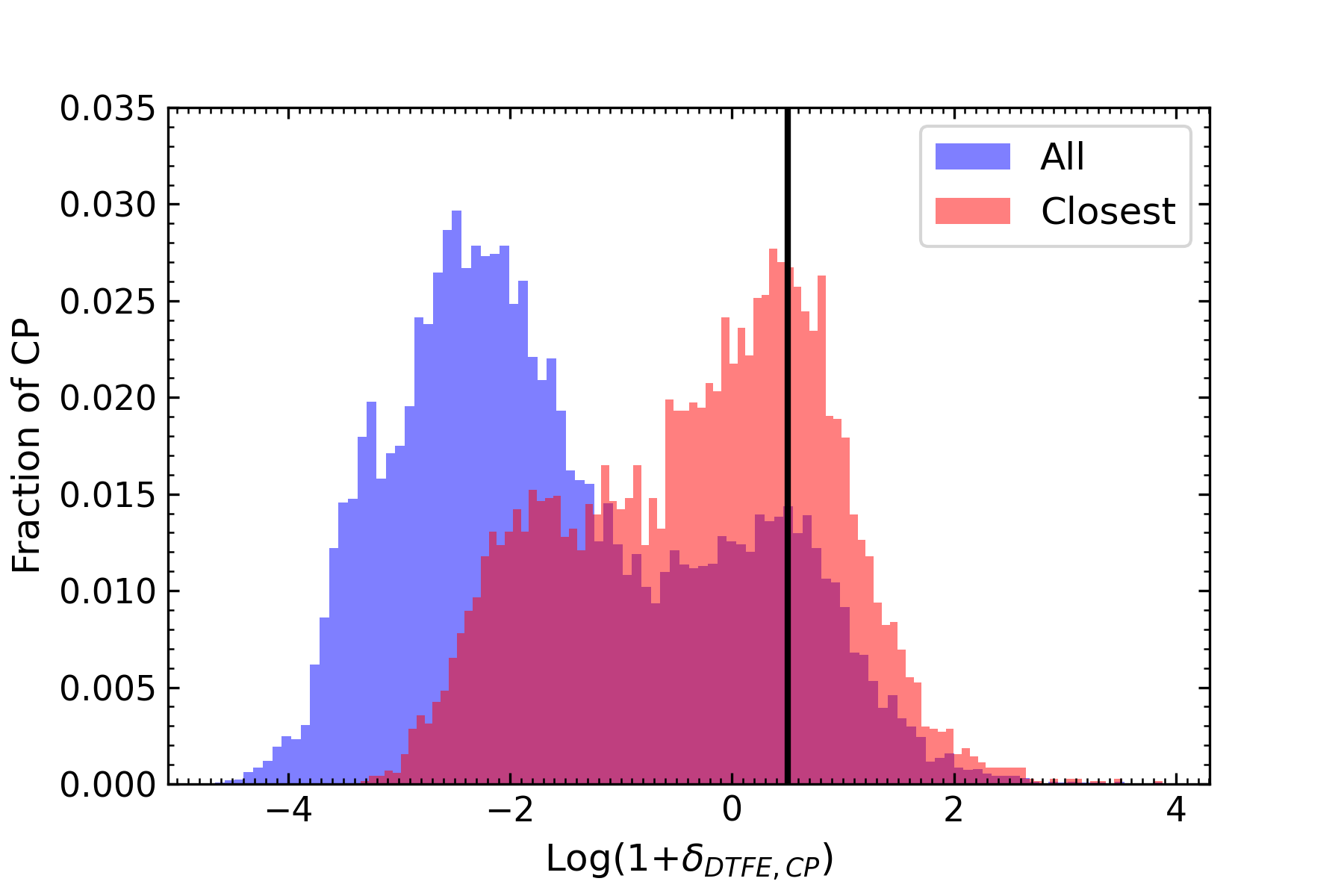}
\caption{Density contrast distribution for all the critical points in the simulation box (blue) and those identified as being closest to at least a subhalo (only maxima and bifurcations considered, in red) when performing the measurement of \dnode. The vertical black line shows the threshold in critical point density contrast adopted.}
\label{cpdensdist}
\end{figure}

Figure \ref{cpdensdist} shows the distribution of $\log(1+\delta_{\mathrm{DTFE, CP}})$ for the critical points, where $\delta_{\mathrm{DTFE, CP}} = \frac{\rho_{\mathrm{DTFE, CP}}-\langle \rho_{\mathrm{DTFE, Gal}} \rangle}{\langle \rho_{\mathrm{DTFE, Gal}} \rangle}$. In this formula, $\rho_{\mathrm{DTFE, CP}}$ is the value of the density computed from the Delaunay tessellation at the position of critical points, while $\rho_{\mathrm{DTFE, Gal}}$ is the density computed at the position of the subhaloes. To measure the average density $\langle \rho_{\mathrm{DTFE, Gal}} \rangle$ we use all the subhaloes in the box. Figure \ref{cpdensdist} shows the density contrast distribution for all the critical points in the simulation box and for those associated with subhaloes, i.e. those that are found to be the closest critical point to at least one subhalo when measuring \dnode.

The density contrast distribution for all critical points is bimodal, highlighting the presence of lower density minima and saddles and higher density bifurcations and maxima. In the case of critical points attached to subhaloes, given that we only have considered maxima and bifurcations, the density contrast distribution is restricted to the high-density tail.

\begin{figure*}
\centering
\includegraphics[trim = 2cm 1cm 1cm 1cm, clip = true, width = \linewidth]{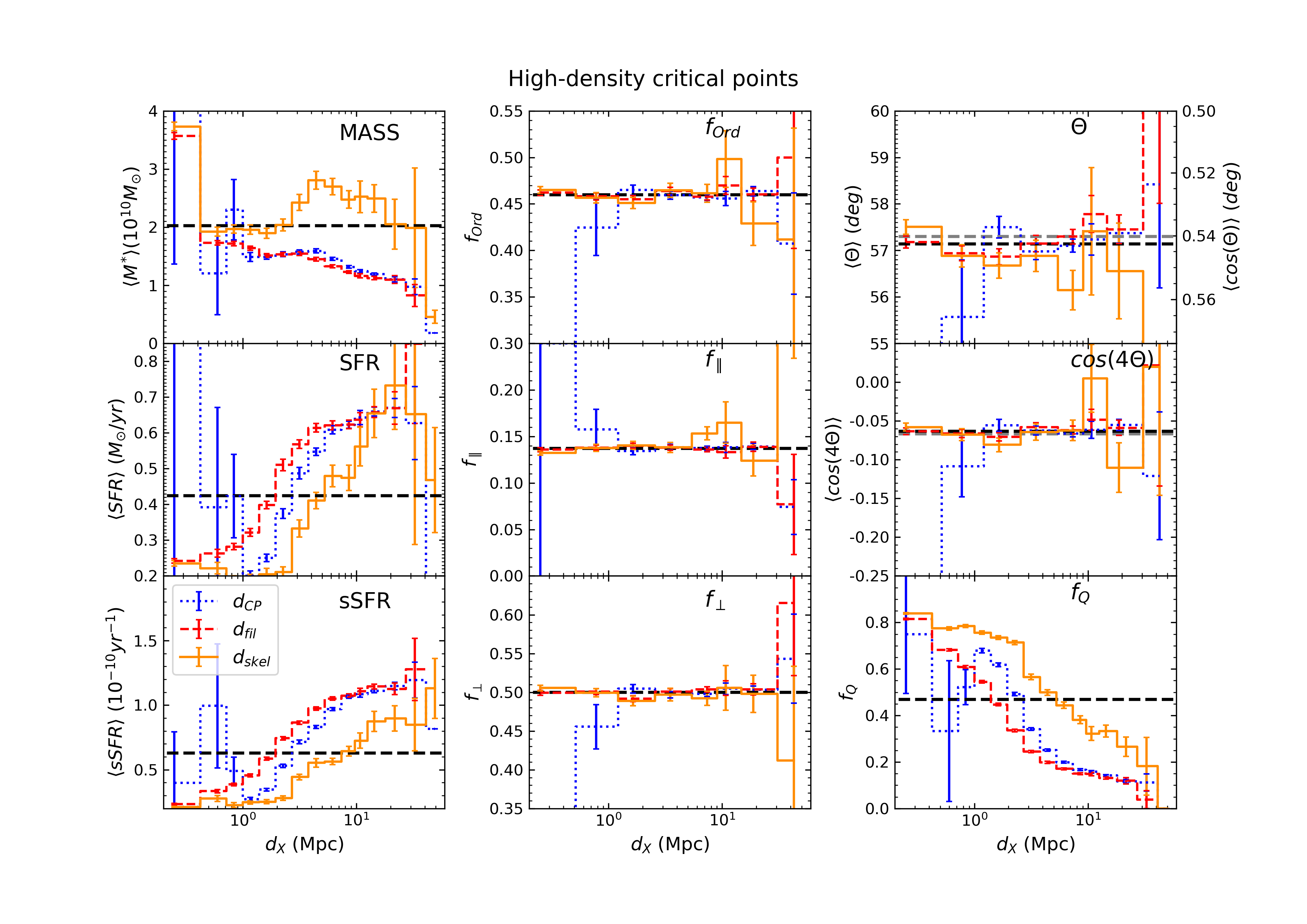}
\caption{Distributions of $\langle M^{\ast} \rangle$, $\langle \mathrm{SFR} \rangle$, $\langle \mathrm{sSFR} \rangle$, $\langle \theta \rangle$, $\langle \cos(4\theta) \rangle$, fraction of quenched galaxies ($f_{\mathrm{Q}}$), fraction of ordered galaxies ($f_{\mathrm{Ord}}$), fraction of parallel galaxies ($f_{\parallel}$), and fraction of perpendicular galaxies ($f_{\perp}$) as a function of the distances from the features of the cosmic web \dfil~ (red dashed line in every panel), \dnode~ (blue dotted line in every panel), and \dskel~ (orange solid line in every panel). In the case of \dnode~ and \dskel, only galaxies outside of filaments (\dfil$\geq 1 \mathrm{Mpc}$) and inside filaments (\dfil$\leq 1 \mathrm{Mpc}$) have been considered, respectively. Error bars on the distributions have been computed through bootstrap resampling. The black dashed line in every panel is the average of the considered quantity in the full simulation box. In the top and middle panel of the central column, the grey line is $\bar{\theta}$ and $\overline{\cos(4\theta)}$, computed given equations \eqref{ptheta} and \eqref{pcos4t} for the total subhalo sample. Only subhaloes whose closer critical point (in terms of \dnode) is above the density contrast threshold of $\log(1+\delta_{\mathrm{DTFE, CP}}) = 0.5$ have been considered. Note that in every panel the first distance bin considered extends all the way to 0 for all distances, however it is cut due to the $x$-axis being in logarithmic scale. Note also that in the $f_{\parallel}$ and $f_{\perp}$ cases the y-axes of the plots cover very different ranges.}
\label{sfrmassangledistclustercut}
\end{figure*}

We adopt a density contrast threshold of $\log(1+\delta_{\mathrm{DTFE, CP}}) = 0.5$ and eliminate from our subhalo sample all galaxies whose closest critical point (considering \dnode) is below the threshold. This leaves a final sample of $102\,064$ subhaloes associated with dense critical points. We re-derive the distributions shown in Figure \ref{sfrmassangledist} using only this subset of galaxies. The new distributions are shown in Figure \ref{sfrmassangledistclustercut}. Aside from the increased noise in the distributions, no major difference from the general case can be seen when only dense critical points close to subhaloes are considered. The only exception is represented by the innermost bin in the \dnode~distribution of spin-related quantities, which now show a much larger fraction of parallel galaxies. However, given for example the large errorbars in the distribution of $f_{\parallel}$ and the very small distance from the critical points considered, this could be due to the reduced number counts in the first bin.

\section{Considering different components to measure the spin}
\label{appendix_spin_components}
\begin{figure}
\centering
\includegraphics[width = \linewidth, trim = 0cm 3cm 0cm 4cm, clip = true]{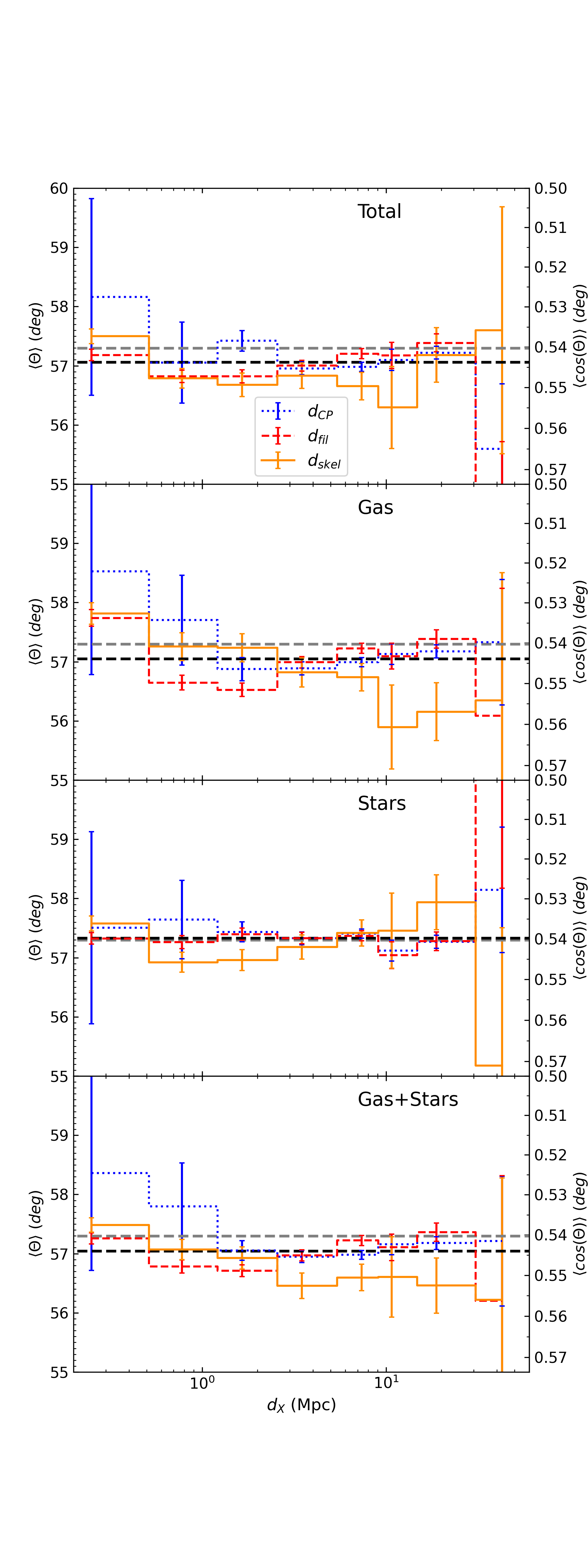}
\caption{Distributions of $\langle \theta \rangle$ as a function of the distances from the features of the cosmic web \dfil~ (red dashed line in every panel), \dnode~ (blue dotted line in every panel), and \dskel~ (orange solid line in every panel). The first panel shows the distributions obtained when all subhalo components are considered (same as top middle panel of Figure \ref{sfrmassangledist}), the subsequent panels refer to the distributions obtained when only the gas component is considered (second panel), only the stellar component (third panel), and when both gas and stars (but not dark matter) are considered (bottom panel). The black dashed line in every panel is the average of the considered quantity in the full simulation box. In the top middle panel, the grey line is $\bar{\theta}$, computed given equation \eqref{ptheta} for the total subhalo sample. Note that in every panel the first distance bin considered extends all the way to 0 for all distances, however it is cut due to the $x$-axis being in logarithmic scale.}
\label{thetadistcomponents}
\end{figure}

As observations are generally able to only probe the baryonic components of galaxies (gas and stars) rather than their dark matter haloes, we explored how the trends with distances from structures for spin-related quantities change when the spin is measured using only the gaseous or stellar components of subhaloes. Figure \ref{thetadistcomponents} shows the distributions of $\langle \theta \rangle$ as a function of \dnode, \dskel, and \dfil, when the spin is measured using all the components (this is the same as top middle panel in Figure \ref{sfrmassangledist}), only the gas, only the stars, or both the gas and stars within each subhalo.

This figure shows that the trend visible in the distributions where the spin of galaxies switches from parallel to perpendicular to the filaments as galaxies flow inside filaments to reach nodes (i.e. where $\langle \theta \rangle$ increases with decreasing \dskel) is essentially due to the gaseous component of galaxies. When the spin is computed using only the gas, the trend with \dskel~ is clearly visible, and it is absent when only the stellar component is considered.

\begin{figure*}
\centering
\includegraphics[width = \linewidth]{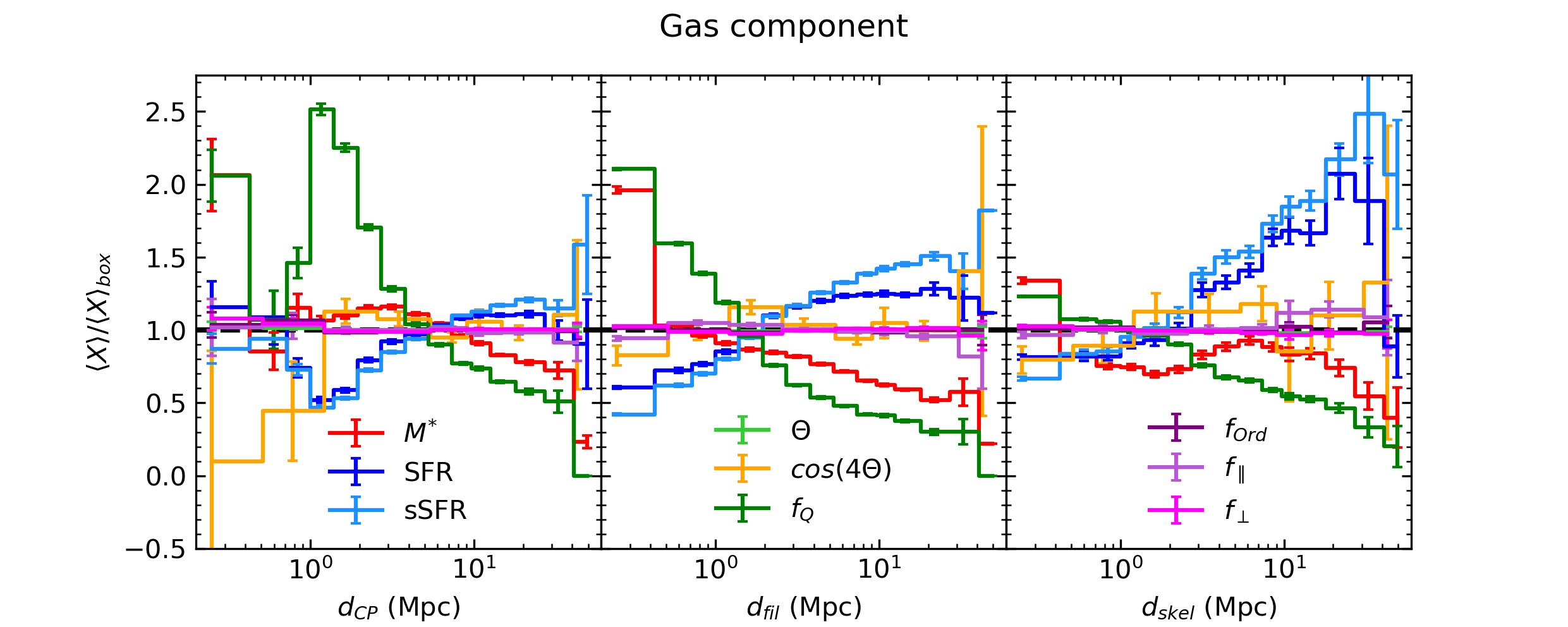}
\caption{Normalised distributions (expressed as $\frac{\langle X \rangle}{\langle X \rangle_{box}}$) with respect to distances \dnode~ (left panel), \dfil~ (middle panel), and \dskel~ (right panel). The quantities considered in each panel ($\langle X \rangle$) are respectively: $\langle M^{\ast} \rangle$ (red), $\langle \mathrm{SFR} \rangle$ (dark blue), $\langle \mathrm{sSFR} \rangle$ (light blue), $\langle \theta \rangle$ (light green), $\langle \cos(4\theta) \rangle$ (orange), $f_{\mathrm{Q}}$ (dark green), $f_{\mathrm{Ord}}$ (dark purple), $f_{\parallel}$ (light purple), and $f_{\perp}$ (magenta). $\langle X \rangle$ indicates the average of the quantity taken including all subhaloes in the box (for \dfil), only those outside filaments (\dnode), and only those inside filaments (\dskel) over the full simulation volume. Only the gas component is used to compute spin-related quantities. Note that in every panel the first distance bin considered extends all the way to 0 for all distances, however it is cut due to the $x$-axis being in logarithmic scale.}
\label{distquantitiessamedistgas}
\end{figure*}

Indeed, when the distributions shown in Figure \ref{distquantitiessamedist} are re-computed with spin-related quantities derived using only the gas component (shown in Figure \ref{distquantitiessamedistgas}), hints of the possibility to use spin-related quantities to trace the cosmic web start to emerge. In particular, in the case of small \dskel~ values, a deviation of $\langle \cos(4\theta) \rangle$ from the average relative to the full box is visible. This deviation is comparable in magnitude to the deviation of the (specific-)SFR. In the case of other spin-related quantities or other distances (\dnode~ and \dfil) no significant variation is visible.

\section{Further checks on the use of the Bullock parameter}
\label{appendix_bullock_r500}
The \citet{Bullock2001} parameter, as currently defined in Section \ref{galpropsspin} makes use of global quantities, meaning that the radius $R$, specific angular momentum $j$ and circular velocity $V_{c}$ entering equation \eqref{bullparam} are computed considering all particles bound to each subhalo, as provided by the IllustrisTNG subhalo catalog. A more proper way to compute the Bullock parameter would be to use virial quantities for the subhaloes. In their work, \citet{Galarraga2019} computed the density profiles for the subhaloes, estimating $R_{200}$ and $M_{200}$ for all the subhaloes in the catalogue. We recompute the specific angular momentum $j_{200}$ using only and all particles within the sphere of radius $R_{200}$ centred on each subhalo. With these quantities we compute $\lambda_{200}$ following equation \eqref{bullparam}. We use the newly computed quantities $j_{200}$ and $\lambda_{200}$ to perform two checks: we re-derive the distributions of spin-related quantities using $j_{200}$ (namely what shown in Figure \ref{sfrmassangledist}) and we divide between high-spin parameter and low-spin parameter galaxies using $\lambda_{200}$ (i.e. the results shown in Figures \ref{sfrmassangledistmasshighspin} and \ref{distquantitiessamedistmassfastrot}).

\begin{figure*}
\centering
\includegraphics[trim = 1cm 1cm 1cm 0cm, clip = true, width = \linewidth]{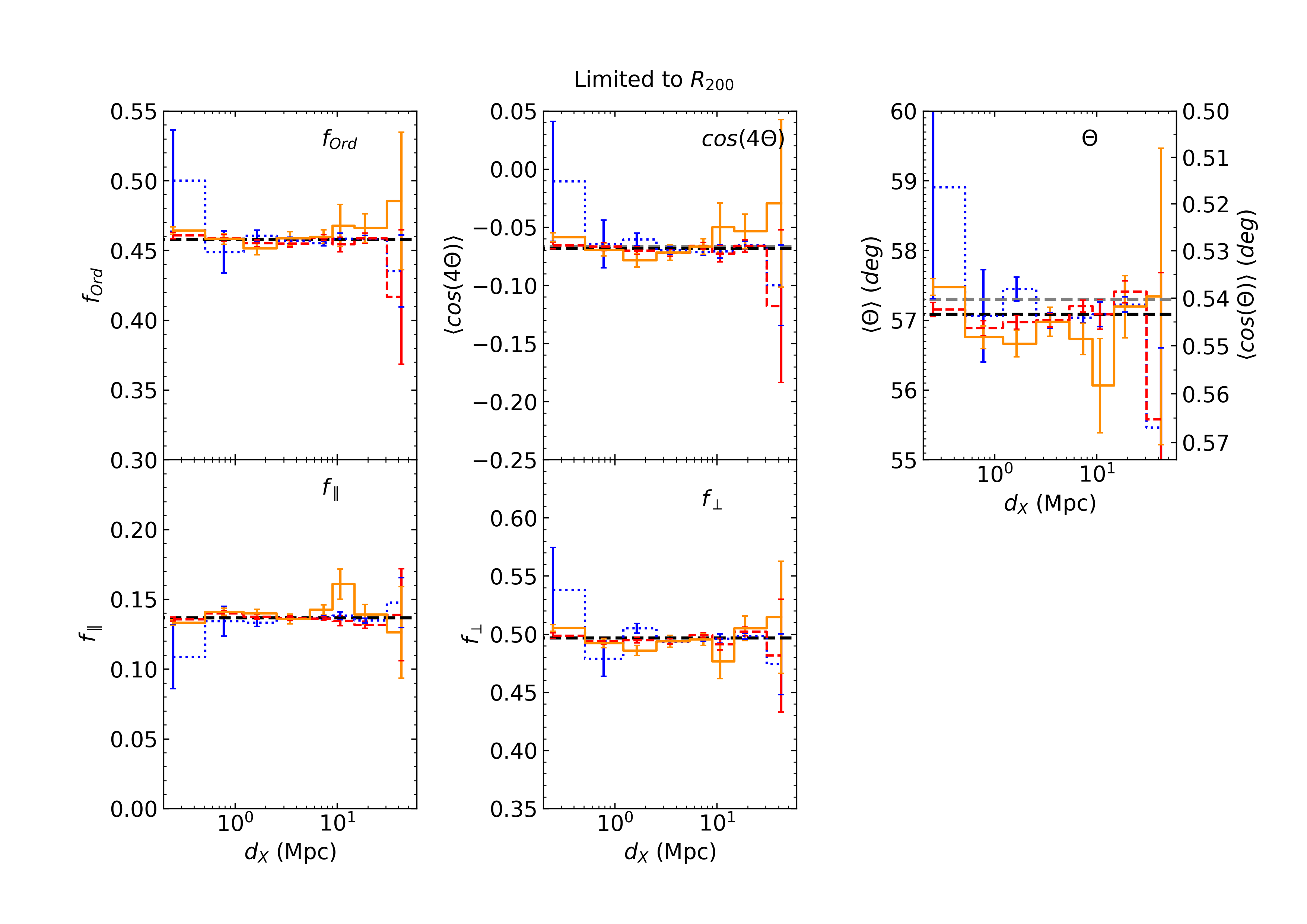}
\caption{Distributions of $\langle \theta \rangle$, $\langle \cos(4\theta) \rangle$, fraction of ordered galaxies ($f_{\mathrm{Ord}}$), fraction of parallel galaxies ($f_{\parallel}$), and fraction of perpendicular galaxies ($f_{\perp}$) as a function of the distances from the features of the cosmic web \dfil~ (red dashed line in every panel), \dnode~ (blue dotted line in every panel), and \dskel~ (orange solid line in every panel). In the case of \dnode~ and \dskel, only galaxies outside of filaments (\dfil$\geq 1 \mathrm{Mpc}$) and inside filaments (\dfil$\leq 1 \mathrm{Mpc}$) have been considered, respectively. Error bars on the distributions have been computed through bootstrap resampling. The black dashed line in every panel is the average of the considered quantity in the full simulation box. In the top left and top middle panels, the grey line is $\bar{\theta}$ and $\overline{\cos(4\theta)}$, computed given equations \eqref{ptheta} and \eqref{pcos4t} for the total subhalo sample. In this figure, the spin $\boldsymbol{j}_{200}$ has been computed using only particles within the $R_{200}$ of every subhalo. Note that in every panel the first distance bin considered extends all the way to 0 for all distances, however it is cut due to the $x$-axis being in logarithmic scale. Note also that in the $f_{\parallel}$ and $f_{\perp}$ cases the y-axes of the plots cover very different ranges.}
\label{thetadistr200}
\end{figure*}

Figure \ref{thetadistr200} shows the distributions of $\langle \theta \rangle$, $\langle \cos(4\theta) \rangle$, fraction of ordered galaxies ($f_{\mathrm{Ord}}$), fraction of parallel galaxies ($f_{\parallel}$), and fraction of perpendicular galaxies ($f_{\perp}$) as a function of \dfil, \dnode, and \dskel for the total galaxy population, when the spin of subhaloes has been computed using only particles within the subhaloes' $R_{200}$. The distributions for spin-related quantities do not show any major difference with case in which the spin has  been computed using all bounded particles in the subhaloes shown in the main text.

\begin{figure*}
\centering
\includegraphics[trim = 2cm 1cm 1cm 0cm, clip = true, width = \linewidth]{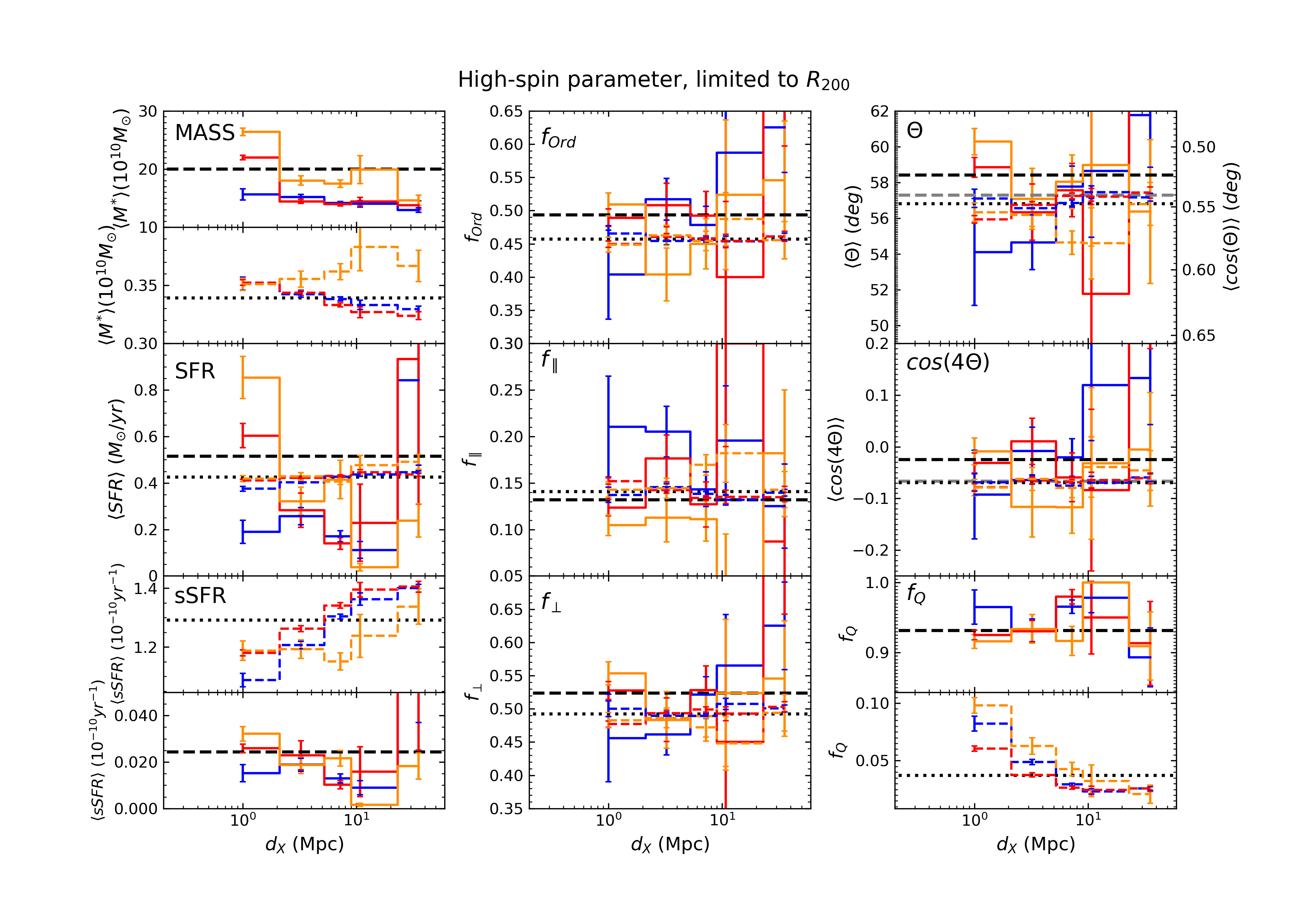}
\caption{Distributions of $\langle M^{\ast} \rangle$, $\langle \mathrm{SFR} \rangle$, $\langle \mathrm{sSFR} \rangle$, $\langle \theta \rangle$, $\langle \cos(4\theta) \rangle$, fraction of quenched galaxies ($f_{\mathrm{Q}}$), fraction of ordered galaxies ($f_{\mathrm{Ord}}$), fraction of parallel galaxies ($f_{\parallel}$), and fraction of perpendicular galaxies ($f_{\perp}$) as a function of the distances from the features of the cosmic web \dfil~ (red line in every panel), \dnode~ (blue line in every panel), and \dskel~ (orange line in every panel) and split between high- and low-mass galaxies. In the case of \dnode~ and \dskel, only galaxies outside of filaments (\dfil$\geq 1 \mathrm{Mpc}$) and inside filaments (\dfil$\leq 1 \mathrm{Mpc}$) have been considered, respectively. Error bars on the distributions have been computed through bootstrap resampling. The black dashed (dotted) line in every panel is the average of the considered quantity in the full simulation box considering only high-mass (low-mass) galaxies. In the top and middle panel of the central column, the grey line is $\bar{\theta}$ and $\overline{\cos(4\theta)}$, computed given equations \eqref{ptheta} and \eqref{pcos4t} for the total subhalo sample. In each panel, solid lines refer to high-mass galaxies and dashed lines to low-mass galaxies. In the case of mass, sSFR, and $f_{\mathrm{Q}}$, the panel has been split in two to take into account the very different ranges on the $y$-axis occupied by the distributions. Only high-spin parameter subhaloes are used to derive the distributions selected using the Bullock parameter computed using only particles within the $R_{200}$ of subhaloes. Note that in every panel the first distance bin considered extends all the way to 0 for all distances, however it is cut due to the $x$-axis being in logarithmic scale. Note also that in the $f_{\parallel}$ and $f_{\perp}$ cases the y-axes of the plots cover very different ranges.}
\label{sfrmassangledistmasshighspinr200}
\end{figure*}

\begin{figure*}
\centering
\includegraphics[width = \linewidth]{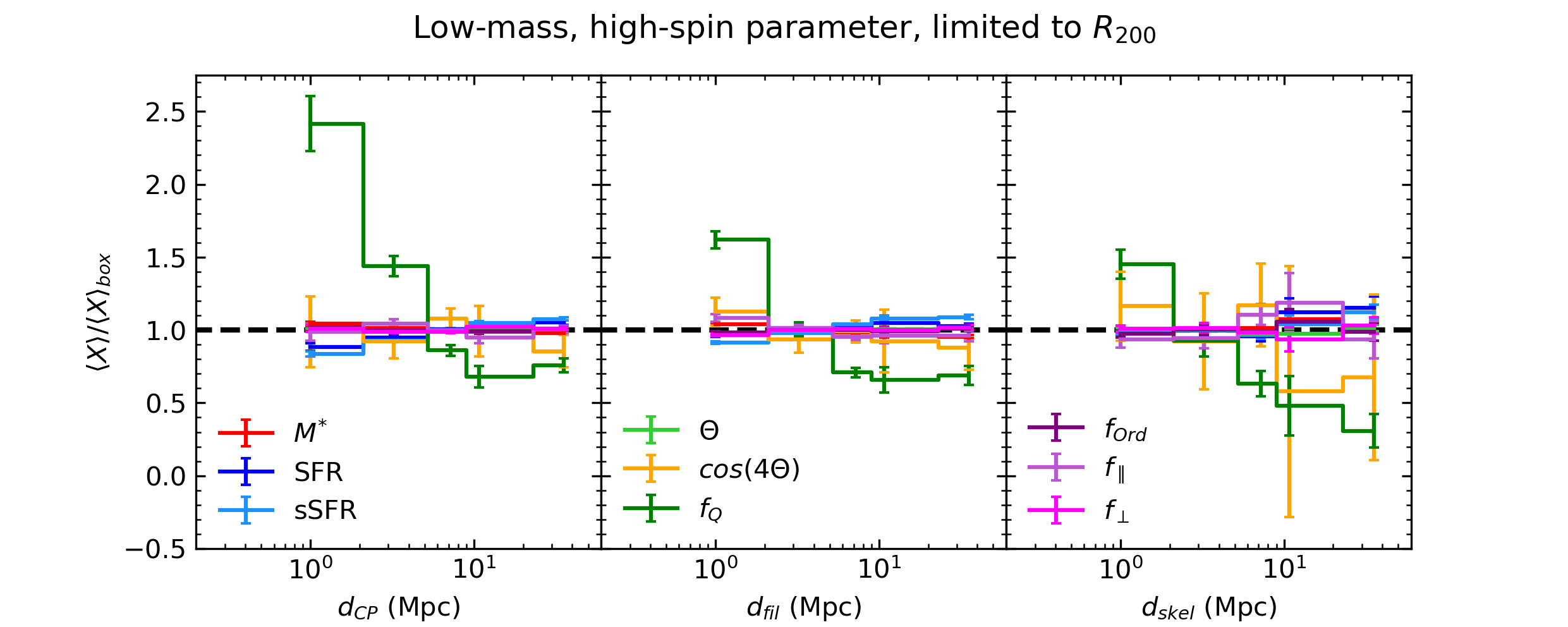}
\caption{Normalised distributions (expressed as $\frac{\langle X \rangle}{\langle X \rangle_{box}}$) with respect to distances \dnode~ (left panel), \dfil~ (middle panel), and \dskel~ (right panel). The quantities considered in each panel ($\langle X \rangle$) are respectively: $\langle M^{\ast} \rangle$ (red), $\langle \mathrm{SFR} \rangle$ (dark blue), $\langle \mathrm{sSFR} \rangle$ (light blue), $\langle \theta \rangle$ (light green), $\langle \cos(4\theta) \rangle$ (orange), $f_{\mathrm{Q}}$ (dark green), $f_{\mathrm{Ord}}$ (dark purple), $f_{\parallel}$ (light purple), and $f_{\perp}$ (magenta). $\langle X \rangle_{box}$ indicates the average of the quantity taken including all high-spin parameter and low-mass subhaloes in the box (for \dfil), only those outside filaments (\dnode), and only those inside filaments (\dskel) over the full simulation volume. Only low-mass, high-spin parameter galaxies are considered, identified by means of the Bullock parameter computed using only particles within $R_{200}$. Note that in every panel the first distance bin considered extends all the way to 0 for all distances, however it is cut due to the $x$-axis being in logarithmic scale.}
\label{distquantitiessamedistmassfastrotr200}
\end{figure*}

Figure \ref{sfrmassangledistmasshighspinr200} shows the distributions of galaxy quantities as a function of the distances to the cosmic web elements for high-spin parameter galaxies (i.e. high $\lambda_{200}$) and separated in high- and low-mass. No major differences can be seen with respect to the case in which the Bullock parameter is computed using all the particles bound to subhaloes. In this case the spin-related quantities have been computed using all the bound particles to the subhaloes, in order to be consistent with what shown in the main text. Particles within $R_{200}$ have been used only to compute the Bullock parameter in order to separate between low-spin parameter and high-spin parameter galaxies. Figure \ref{distquantitiessamedistmassfastrotr200} shows an analogous of Figure \ref{distquantitiessamedistmassfastrot} only for high-spin parameter galaxies, selected using $\lambda_{200}$. Again, no difference with respect to the general case in which all particles bound to a subhalo are used to measure the Bullock parameter can be identified.

\end{document}